\documentclass[twocolumn,showpacs,aps,prd,superscriptaddress]{revtex4}

\usepackage{xspace}
\usepackage{graphicx,rotating}
\usepackage{dcolumn}
\usepackage{amsmath}
\usepackage{epsfig}
\usepackage{dcolumn}

\RequirePackage{xspace}

\usepackage{relsize}
\def\babar{\mbox{\slshape B\kern-0.1em{\smaller A}\kern-0.1em
    B\kern-0.1em{\smaller A\kern-0.2em R}}}

\def\b     {\ensuremath{b}\xspace}

\def\piz   {\ensuremath{\pi^0}\xspace}

\def\pip   {\ensuremath{\pi^+}\xspace}
\def\pim   {\ensuremath{\pi^-}\xspace}

\def\kaon  {\ensuremath{K}\xspace}
\def\Kbar  {\kern 0.2em\overline{\kern -0.2em K}{}\xspace}

\def\Kz    {\ensuremath{K^0}\xspace}
\def\Kzb   {\ensuremath{\Kbar^0}\xspace}
\def\KzKzb {\ensuremath{\Kz \kern -0.16em \Kzb}\xspace}
\def\Kp    {\ensuremath{K^+}\xspace}
\def\Km    {\ensuremath{K^-}\xspace}

\def\KpKm  {\ensuremath{\Kp \kern -0.16em \Km}\xspace}
\def\KS    {\ensuremath{K^0_{\scriptscriptstyle S}}\xspace} 
\def\KL    {\ensuremath{K^0_{\scriptscriptstyle L}}\xspace}

\def\Kstar   {\ensuremath{K^*}\xspace}

\def\Dbar    {\kern 0.2em\overline{\kern -0.2em D}{}\xspace}

\def\Dz      {\ensuremath{D^0}\xspace}
\def\Dzb     {\ensuremath{\Dbar^0}\xspace}
\def\DzDzb   {\ensuremath{\Dz {\kern -0.16em \Dzb}}\xspace}
\def\Dp      {\ensuremath{D^+}\xspace}
\def\Dm      {\ensuremath{D^-}\xspace}

\def\DpDm    {\ensuremath{\Dp {\kern -0.16em \Dm}}\xspace}

\def\B       {\ensuremath{B}\xspace}
\def\Bbar    {\kern 0.18em\overline{\kern -0.18em B}{}\xspace}

\def\BB      {\ensuremath{B\Bbar}\xspace} 
\def\Bz      {\ensuremath{B^0}\xspace}
\def\Bzb     {\ensuremath{\Bbar^0}\xspace}
\def\BzBzb   {\ensuremath{\Bz {\kern -0.16em \Bzb}}\xspace}
\def\Bu      {\ensuremath{B^+}\xspace}
\def\Bub     {\ensuremath{B^-}\xspace}

\def\Bm      {\ensuremath{\Bub}\xspace}

\def\BpBm    {\ensuremath{\Bu {\kern -0.16em \Bub}}\xspace}

\mathchardef\Upsilon="7107
\def\Y#1S{\ensuremath{\Upsilon{(#1S)}}\xspace}% no space before {...}!

\def\FourS {\Y4S}

\mathchardef\Deltares="7101
\mathchardef\Xi="7104
\mathchardef\Lambda="7103
\mathchardef\Sigma="7106
\mathchardef\Omega="710A

\def\Deltabar{\kern 0.25em\overline{\kern -0.25em \Deltares}{}\xspace}
\def\Lbar{\kern 0.2em\overline{\kern -0.2em\Lambda\kern 0.05em}\kern-0.05em{}\xspace}
\def\Sigbar{\kern 0.2em\overline{\kern -0.2em \Sigma}{}\xspace}
\def\Xibar{\kern 0.2em\overline{\kern -0.2em \Xi}{}\xspace}
\def\Obar{\kern 0.2em\overline{\kern -0.2em \Omega}{}\xspace}
\def\Nbar{\kern 0.2em\overline{\kern -0.2em N}{}\xspace}
\def\Xb{\kern 0.2em\overline{\kern -0.2em X}{}\xspace}

\def\mes        {\mbox{$m_{\rm ES}$}\xspace}

\def\DeltaE     {\mbox{$\Delta E$}\xspace}

\newcommand{\tev}{\ensuremath{\mathrm{\,Te\kern -0.1em V}}\xspace}
\newcommand{\gev}{\ensuremath{\mathrm{\,Ge\kern -0.1em V}}\xspace}
\newcommand{\mev}{\ensuremath{\mathrm{\,Me\kern -0.1em V}}\xspace}
\newcommand{\kev}{\ensuremath{\mathrm{\,ke\kern -0.1em V}}\xspace}
\newcommand{\ev}{\ensuremath{\mathrm{\,e\kern -0.1em V}}\xspace}
\newcommand{\gevc}{\ensuremath{{\mathrm{\,Ge\kern -0.1em V\!/}c}}\xspace}
\newcommand{\mevc}{\ensuremath{{\mathrm{\,Me\kern -0.1em V\!/}c}}\xspace}
\newcommand{\gevcc}{\ensuremath{{\mathrm{\,Ge\kern -0.1em V\!/}c^2}}\xspace}
\newcommand{\mevcc}{\ensuremath{{\mathrm{\,Me\kern -0.1em V\!/}c^2}}\xspace}
\def\invfb   {\ensuremath{\mbox{\,fb}^{-1}}\xspace}

\def\mus  {\ensuremath{\rm \,\mus}\xspace}

\def\mus        {\ensuremath{\,\mu{\rm s}}\xspace}    %% microsecond
\def\to                 {\ensuremath{\rightarrow}\xspace}

\def\pep2{PEP-II}

\newcommand{\dedx}{\ensuremath{\mathrm{d}\hspace{-0.1em}E/\mathrm{d}x}\xspace}

\def\gsim{{~\raise.15em\hbox{$>$}\kern-.85em
          \lower.35em\hbox{$\sim$}~}\xspace}
\def\lsim{{~\raise.15em\hbox{$<$}\kern-.85em
          \lower.35em\hbox{$\sim$}~}\xspace}

\def\Vub  {\ensuremath{|V_{ub}|}\xspace}

\xspace

\newcommand{\epjBase}        {Eur.\ Phys.\ Jour.\xspace}

\newcommand{\jprBase}        {Phys.\ Rev.\xspace}
\newcommand{\jplBase}        {Phys.\ Lett.\xspace}
\newcommand{\nimBaseA}       {Nucl.\ Instr.\ Meth.\xspace}

\newcommand{\nimBaseC}       {Nucl.\ Instr.\ and Methods\xspace}

\newcommand{\npBase}         {Nucl.\ Phys.\xspace}
\newcommand{\zpBase}         {Z.\ Phys.\xspace}

\newcommand{\epjc}      [1]  {\epjBase\ C~{\bf #1}}

\newcommand{\mpl}       [1]  {{Mod.\ Phys.\ Lett.\ {\bf #1}}}

\newcommand{\nim}       [1]  {\nimBaseC~{\bf #1}}
\newcommand{\nima}      [1]  {\nimBaseA~A~{\bf #1}}

\newcommand{\npb}       [1]  {\npBase\ B~{\bf #1}}

\newcommand{\npbps}     [1]  {{Nucl.\ Phys.\ B~Proc.\ Suppl.\ {\bf #1}}}

\newcommand{\plb}       [1]  {\jplBase\ B~{\bf #1}}

\newcommand{\pr}        [1]  {\jprBase\ {\bf #1}}

\newcommand{\progtp}    [1]  {{Prog.\ Th.\ Phys.\ {\bf #1}}}

\newcommand{\zpc}       [1]  {\zpBase\ C~{\bf #1}}

\def\jetset74   {\mbox{\tt Jetset \hspace{-0.5em}7.\hspace{-0.2em}4}\xspace}
 \def\er #1 #2 { $#1 \pm #2$ }
 \def\bra #1 #2 #3 #4 { $#1 ^{+#2} _{-#3} \pm #4 $ }

\newcommand{\BABARPubYear}     {??}
\newcommand{\BABARPubNumber}  {???}

\newcommand{\SLACPubNumber} {?????}

\begin{document}
\preprint{\babar-PUB-\BABARPubYear/\BABARPubNumber} 
\preprint{SLAC-PUB-\SLACPubNumber} 
%
%% Needed in final document
\begin{flushleft}
\babar-PUB-05/37\\
SLAC-PUB-11365\\
hep-ex/0508004\\[10mm]
\end{flushleft}
\title{
{\large \bf \boldmath
Measurements of the $\B \to X_s \gamma$ Branching Fraction and Photon Spectrum from a Sum of 
Exclusive Final States} 
}
%
% author list; 
%\input authors_jun2005.tex
%% author list as of 02-Jun-2005 (635 authors)
%
\author{B.~Aubert}
\author{R.~Barate}
\author{D.~Boutigny}
\author{F.~Couderc}
\author{Y.~Karyotakis}
\author{J.~P.~Lees}
\author{V.~Poireau}
\author{V.~Tisserand}
\author{A.~Zghiche}
\affiliation{Laboratoire de Physique des Particules, F-74941 Annecy-le-Vieux, France }
\author{E.~Grauges}
\affiliation{IFAE, Universitat Autonoma de Barcelona, E-08193 Bellaterra, Barcelona, Spain }
% ...
\collaboration{The \babar\ Collaboration}
\noaffiliation
\date{\today}% It is always \today, today, but you may specify any date with \date.
\begin{abstract}
Using 88.9 million \BB\ events collected by the \babar\ detector at the \FourS, we measure 
the branching fraction for the radiative penguin process $\B\to X_s\gamma$ 
from the sum of 38 exclusive final states. The inclusive branching fraction 
above a minimum photon energy $E_{\gamma}>$1.9~\gev\ is
${\cal B}(\b\to s\gamma)=(3.27\pm 0.18(stat.)^{+0.55}_{-0.40}(syst.)^{+0.04}_{-0.09}(theory))\times 10^{-4}$.
We also measure the isospin asymmetry between 
$\Bm\to X_{s\bar{u}}\gamma$ and $\Bzb \to X_{s\bar{d}}\gamma$
to be $\Delta_{0-} = -0.006 \pm 0.058(stat.)\pm 0.009(syst.)\pm 0.024(\Bzb\ /\Bm)$.
The photon energy spectrum is measured in the $B$ rest frame, 
from which moments are derived for different values of the minimum photon energy. 
We present fits to the photon spectrum and moments which give the heavy--quark parameters 
$m_\b$ and $\mu_{\pi}^2$.
The fitted parameters are consistent with those obtained from semileptonic $\B\to X_c\ell\nu$ decays, and 
are useful inputs for the extraction of \Vub\ from measurements of semileptonic $\B\to X_u\ell\nu$ decays.

\end{abstract}
\maketitle
\section{Introduction} \label{sec:intro}

Radiative decays involving the flavor--changing neutral current transition 
$b\to s$ are described in the Standard Model primarily by 
a one--loop radiative penguin diagram containing a top quark and a $W$ boson. 
Calculations of this rate in the Standard Model have 
now been completed to next-to-leading order in $\alpha_s$ (NLO), 
with a predicted branching fraction 
${\cal B}(\b\to s\gamma)=(3.57\pm 0.30)\times 10^{-4}$
for $E_\gamma>1.6\,\gev$~\cite{TheoryBF1, TheoryBF2, TheoryBF3}, which is consistent with the current 
experimental world average ${\cal B}(\b\to s\gamma)=(3.52^{+0.30}_{-0.28})\times 10^{-4}$~\cite{HFAG}.
Calculations of next-to-next-to-leading order (NNLO) corrections are in progress~\cite{NNLO}.
Additional contributions to the loop from new physics, 
e.g. a charged Higgs boson, could change the $b\to s\gamma$ 
rate~\cite{NewPhysics1,NewPhysics2,NewPhysics3,KaganNeubert,NewPhysics4}.

The photon energy spectrum in $B\to X_{s}\gamma$ provides access to the distribution 
function of the $b$ quark inside the \B\ meson~\cite{distributionfunction}. The knowledge of this shape
function is a crucial input in the extraction of \Vub\ from inclusive 
semileptonic $\B\to X_u\ell\nu$ 
measurements~\cite{ShapeFunction1,ShapeFunction2,ShapeFunction3,ShapeFunction4,ShapeFunction5,ShapeFunction6}. 
We fit the spectrum to two recent predictions, one using a combination of the operator 
product expansion (OPE) coupled to soft 
collinear effective theory~\cite{TheoryBF3,ShapeFunction4,ShapeFunction5,ShapeFunction6,N1,N2} 
in the shape function mass scheme,
and the other using a full OPE approach in the kinetic mass scheme~\cite{BBU}.

We also present measurements in the \B\ rest frame of the first, second and third
moments of the photon energy spectrum for five different minimum energies, 
$E_\gamma>1.90$, 2.00, 2.09, 2.18 and $2.26\,\gev$. The heavy quark parameters $m_\b$ 
and $\mu_{\pi}^2$, which describe the effective $\b$ quark mass and kinetic energy 
inside the \B\ meson, can be determined either from 
fits to these moments~\cite{BBU}, or from the fits to the spectrum.
We compare the fitted parameters with 
those obtained from the lepton energy and hadronic mass moments measured in
semileptonic $\B\to X_c\ell\nu$ decays~\cite{Vcb}.
Previous measurements of the inclusive branching fraction have used two different methods.
In the fully inclusive method~\cite{cleo,aleph,belleinclbsg} the photon energy spectrum is measured without 
reconstructing the $X_s$ system, and backgrounds are suppressed using information from the rest of the event.
When measuring the $E_\gamma$ spectrum inclusively at the \FourS\ the
shape of the spectrum has a large contribution from 
the 50\,\mev\ calorimeter energy resolution, 
and from the motion of the \B\ meson in the \FourS\ rest frame.

The semi-inclusive method~\cite{cleo,bellesemiincl,babaracp}  
uses a sum of exclusive final states in which possible $X_s$ systems are 
combined with the photon, and kinematic constraints of \FourS\ production
are used to suppress backgrounds.  
We have chosen this method and made several improvements over the previous 
analyses. The number of final states of the $X_s$ system has been increased 
to 38 by the inclusion of states with two $\piz$s, $\eta$ mesons, and 
three kaons, and the $X_s$ mass range has been increased to $0.6-2.8\,\gev$. 
Candidates with correctly reconstructed $X_s$ systems are treated as signal, whereas 
incorrectly reconstructed systems, referred to as ``cross-feed'', are treated as background. 
This method allows us to make a measurement of 
the branching fraction as a function of the hadronic mass, $M(X_s)$. 
The $M(X_s)$ spectrum is converted into a high resolution photon energy spectrum 
using the kinematic relationship for the decay of a \B\ meson of mass $M_\B$:
\begin{equation}
\label{eq:egamma}
   E_\gamma = {{M_\B^2 - M(X_s)^2}\over{2M_\B}}.
\end{equation}
where $E_\gamma$ is the photon energy in the \B\ rest frame 
which has a resolution of 1--5\,\mev. 

\section{Detector and Data} \label{sec:detector}

The results presented in this paper are based on data collected
with the \babar\ detector~\cite{Detector}
at the PEP-II asymmetric-energy $e^+e^-$ collider
located at the Stanford Linear Accelerator Center.  The data sample
has an integrated luminosity of 81.9\,\invfb, 
corresponding to 88.9 million \BB\ pairs
recorded at the \FourS\ resonance
(``on-peak'', center-of-mass energy $\sqrt{s}=10.58\,\gev$).
An additional 9.6\,\invfb\ were recorded about 40\,\mev\ below
this energy (``off-peak''), for the study of continuum backgrounds in
which a light or charm quark pair is produced.

The asymmetric beam configuration in the laboratory frame
provides a boost of $\beta\gamma = 0.56$ to the \FourS.
This results in the high energy photons from $b\to s\gamma$ decays 
having energies between 1.5 and 4.5\,\gev\ in the laboratory frame.\
Photons are detected and their energies measured by a CsI(Tl) 
electromagnetic calorimeter (EMC).  
The energy scale of the calorimeter crystals is determined by 
radioactive source and Bhabha scattering calibrations, and 
the energy leakage of photon showers is corrected using $\piz$ decays.
The photon energy resolution is measured 
with symmetric $\piz$ and $\eta$ decays to be
$\sigma_{E}/E = \left\{2.3 / E(\gev)^{1/4} \oplus 1.4 \right\}\%$, 
where the terms are added in quadrature. 
The measured $\piz$ mass resolution is between 6 and 7\,\mev\
for momenta below 1\,\gev\ in the laboratory frame.

Charged particles are detected and their momenta measured by the
combination of a silicon vertex tracker, consisting of five layers
of double-sided detectors, and a 40-layer central drift chamber,
both operating in the 1.5\,T magnetic field of a solenoid.  The transverse 
momentum resolution for the tracking system is 
$\sigma_{p_T}/p_T=0.0013p_T\oplus 0.0045$, where $p_T$ is measured in $\gev$.  

Charged particle identification is provided by the average 
energy loss (\dedx) in the tracking devices and
by an internally reflecting ring-imaging Cherenkov detector (DIRC).  
The $\dedx$ resolution from the drift chamber is typically $7.5\%$ for pions.  
The Cherenkov angle resolution of the DIRC is measured to be 2.4\,mrad, which
provides more than $3\sigma$ separation between charged kaons and pions up to a
momentum of $3\,\gev$. 

\section{$X_S$ Signal Model and Backgrounds} \label{sec:sigmodel}

The $B\to X_{s}\gamma$ signal includes resonant and non-resonant $X_{s}$ states, 
but S-wave states are forbidden by angular momentum conservation.
It is known experimentally that the mass region $M(X_s)<1.1\,\gev$
is dominated by the $\Kstar(892)$ resonance~\cite{Kstar}.
In the higher-mass region there is evidence for the 
$K_1(1270)$ and $K_2^{*}(1430)$ resonances~\cite{Kstarstar}, but these only 
account for about 16\,\% of the inclusive rate. Theoretical predictions 
for exclusive decays to higher $\Kstar$ resonances also account for 
less than half of the inclusive rate~\cite{KstarstarTheory}.
 
The sum of many broad resonances can be modeled by an inclusive distribution.
In designing our analysis we have used an inclusive 
calculation from Kagan and Neubert~\cite{KaganNeubert}. We 
follow their prescription and replace the inclusive model 
in the region $M(X_s)<1.1\,\gev$ with an equivalent amount of exclusive $\B\to \Kstar(892)\gamma$.
The Kagan and Neubert calculation has two empirical parameters, 
$m_b$ and $\lambda_1$, which are related to the mean and width of the spectrum 
(they are similar to the OPE parameters $m_\b$ and $-\mu_{\pi}^2$).
The default parameters were originally chosen to be $m_\b=4.65\,\gev$ and $\lambda_1 = -0.52\,\gev^2$,  
but eventually we fit for these parameters using our measured spectrum.   
In the inclusive region $M(X_s)=1.1-2.8\,\gev$ 
the fragmentation of the $X_s$ system into hadrons is simulated
using JETSET~\cite{JETSET}. The response of the detector 
is modeled using GEANT4~\cite{GEANT4}.

Most of the background in this analysis arises from continuum production of a high 
energy photon, either by initial state radiation, or from the decays of 
$\piz$ and $\eta$ mesons produced in light--quark jets. 
We combine event shape information into a neural network and use the output 
to remove most of this background. The $\piz$ and $\eta$ contributions are further suppressed
by vetoes on combinations of the high energy photon with another photon 
in the event which have a mass consistent with a $\piz$ or $\eta$.

Backgrounds from hadronic $b\to c$ decays are important for $M(X_s)>1.8\,\gev$.
There are two contributions to this background: combinatorial  
final states containing particles from both \B\ decays, 
and incomplete final states where the particles all come from the 
decay of one \B, but one or more of the decay products is missing. 
In the case when only one low energy photon is missing from a \B\ decay to a final state 
containing a $\piz$, e.g. $\B\to D^{(*)}\rho^-$, the events
tend to peak in the signal region, since only a small amount of energy is missing from the final state. 
At high hadronic masses this background becomes comparable to the expected signal yield.

Backgrounds from charmless hadronic \B\ decays give a small contribution over the whole
$X_s$ mass range, but they include a poorly understood component from $\B\to X_s\piz$ decays
which can peak in the signal region.
These backgrounds are modeled as a sum of the measured
and yet unmeasured charmless decay modes.
There is also a cross-feed background from mis-reconstructed 
$\B\to X_s\gamma$ decays which is discussed in detail in 
Section~\ref{sec:efficiency}. The requirement of a positively identified kaon removes
$B\to X_d\gamma$ decays.

\section{\B\ Meson Reconstruction} \label{sec:selection}

We reconstruct the $X_s$ states in 38 decay modes and their charge conjugates. This
includes 22 final states with a kaon and one to four pions, 
where at most two of the pions are $\piz$s, 10 states with a kaon, an $\eta$ 
and up to two pions, 
and 6 states with three kaons plus at most one pion.
The kaons can be either $\Km$ or $\KS$.  A full list of the modes 
can be found in Table~\ref{tab:fragsubset}. 
According to our signal model these modes represent 55\,\% 
of the total inclusive rate in the region $M(X_s)=1.1-2.8\,\gev$. 
The $X_s$ modes that we do not reconstruct are referred
to as ``missing fractions''.

Neutral kaons are reconstructed as $\KS\to\pip\pim$ candidates with an invariant mass
within 9\,\mev\ of the nominal $\KS$ mass~\cite{PDG}, and a transverse flight distance 
greater than 2\,mm from the
primary event vertex. We do not reconstruct $\KS\to\piz\piz$ because of its low efficiency, and we do not 
reconstruct $\KL$ because we cannot directly measure its energy.
Charged kaons are identified using information from the DIRC and
the tracking devices. The remaining tracks are considered to be from charged pions.
Both charged and neutral kaons are required to have momenta $>$ 0.7\,\gev\ in the laboratory frame. 
Above this threshold the rate for charged pions to be mis-identified as kaons is $<$ 2\,\%.

Neutral pions are reconstructed from pairs of photons, each with an energy $>$ 30\,\mev.
For $\piz$ candidates in the mass interval 117 and 150\,\mev, a fit is performed to
improve the momentum resolution. To reject combinatorial background, charged and neutral pions 
are required to have momenta $>$ 0.5, 0.3 or 0.2\,\gev\ in the laboratory frame for states with 
1, 2 or $\ge 3$ pions, respectively.

The $\eta$ mesons are reconstructed from pairs of photons with energies $>$ 50\,\mev.  
For $\eta$ candidates in the mass interval 520 and 580\,\mev, a fit is performed to
improve the momentum resolution. The $\eta$ mesons are required to have momenta $>$~0.7\,\gev\ in the 
laboratory frame.
We do not explicitly reconstruct the modes $\eta\to\pip\pim\piz$ and $\eta\to\piz\piz\piz$, 
but the former decays are included in the final states with a kaon and $\ge 3$ pions. 

The reconstructed $X_s$ system is combined with a high--energy photon to form a \B\ meson. 
The photon is detected as an isolated energy cluster in the calorimeter, with
shape consistent with a single photon, and energy $E_{\gamma}^*>1.8$\,\gev\ in the $e^+e^-$ 
center--of--mass (CM) frame.
A veto is applied to high energy photons that, combined with another photon, form either a $\piz$ within
the mass range 117--150\,\mev\ or an $\eta$ within the mass range 524--566\,\mev.
In the higher-mass region $M(X_s)>2.0\,\gev$, we improve the rejection of $\B\to D^{(*)}\rho^-$ background 
by opening up the $\piz$ mass window to 106--162\,\mev\, and applying a veto if the \piz\ forms 
a $\rho^-$ meson in the mass range 400-1200\,\mev\ when combined with a charged pion.  

We remove 85\,\% of the continuum background by selections on
the angle, $\theta_T^*$, between the thrust axis of the \B\ meson candidate and the thrust axis of all
the other particles of the event, requiring $|\cos\theta_T^*| < 0.80$, and the angle, $\theta_\B^*$, 
between the \B\ candidate and the beam axis, requiring $|\cos\theta_\B^*|<0.80$, both defined in the $e^+e^-$
CM system. We then use a neural network to combine information from a
set of event-shape variables. The inputs to the neural net include $\theta_T^*$, $\theta_\B^*$, 
$R2$, the ratio of the second to zeroth Fox-Wolfram moments~\cite{foxwolfram}, and $R2'$ and  $\theta_T'$, which are defined 
in the primed frame obtained by removing the high energy photon and boosting the rest of the event into 
its CM frame. These last two variables discriminate against background from initial state radiation. 
The inputs to the neural net also include a set of 18 energy flow cones each covering an angle of $10^{\circ}$ 
about the reconstructed $B$ direction in the $e^+e^-$ CM system. 
The neural network selection is tightened above $M(X_s)=1.1\,\gev$, and again 
above $M(X_s)=2.0\,\gev$ and $M(X_s)=2.4\,\gev$, to take account of the increasing background as a 
function of hadronic mass.  

The identification of $\B\to X_s\gamma$ decays makes use of two kinematic variables:
the beam--energy substituted mass, $\mes= \sqrt{(\sqrt{s}/2)^2- p^{*2}_\B}$,
and the difference between the measured and expected energies of the \B\ candidate,
$\DeltaE = E_\B^* - (\sqrt{s}/2)$, where $E_\B^*$ and $p_\B^*$ are the energy and
momentum of the \B\ candidate in the CM frame, and
$\sqrt{s}$ is the total CM energy.
When calculating $\mes$, the value of $p^*_\B$ is corrected for the tail of the
high energy photon response function of the EMC by scaling the measured $E_{\gamma}^*$ 
to the value that gives $\DeltaE = 0$, the value expected for true signal.

Within an initial selection $\mes>5.22\,\gev$ and $|\DeltaE|<0.40\,\gev$ 
we reconstruct two candidates per event on average.
In events where more than one \B\ candidate is reconstructed
we select the best candidate using the smallest value of $|\DeltaE|$. 
This technique is $>$ 90\,\%\ efficient when the true $\B\to X_s \gamma$
decay is among the reconstructed candidates. 
The $|\DeltaE|$ distribution has a resolution of about 0.05\,\gev\ with a radiative tail on the low side. 
For the best candidate we require $|\DeltaE|<0.10\,\gev$ for $M(X_s)<2.0\,\gev$, and tighten this  
to $|\DeltaE|<0.08\,\gev$ for $M(X_s)=2.0-2.4\,\gev$ and $|\DeltaE|<0.07\,\gev$ for $M(X_s)=2.4-2.8\,\gev$. 
These selections are optimized to give the best statistical significance for the signal in each 
$X_s$ region. We then fit the $\mes$ distribution between 5.22 and 5.29\,\gev\ to extract the signal yield.

\begin{figure}[htbp]
\begin{center}
 \begin{tabular}{c}
   \mbox{\includegraphics[width=3.0in]{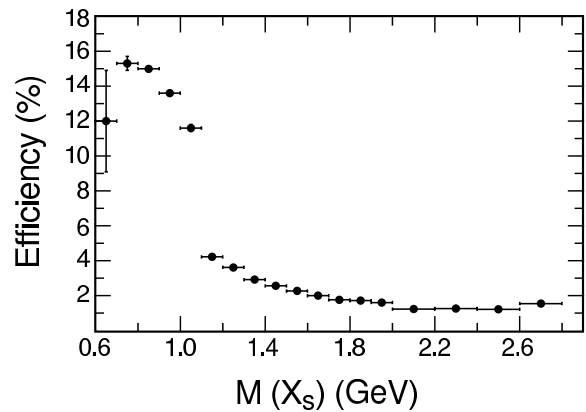}}\\
 \end{tabular}
\end{center}    
\caption{Efficiency for correctly reconstructing a signal event in one of the 38 final states 
as a function of hadronic mass. 
Note that this efficiency does not include the missing fractions
of $B\to X_{s}\gamma$ final states.} 
\label{fig:eff_mass}
\end{figure}

\section{Signal Efficiency and Cross-feed} \label{sec:efficiency}

The signal efficiency is determined from generated Monte Carlo events which are produced and correctly 
reconstructed in one of the 38 final states. It does not include the missing 
$B\to X_{s}\gamma$ final states which are discussed in Section~\ref{sec:missing}.
We generate equal numbers of \Bm\ and \Bzb\ decays and assume isospin 
symmetry. The production of \Km\ and \Kz\ is equal, and the branching fraction 
for the $\KS\to\pip\pim$ decay is included in the Monte Carlo generator. 
For the isospin asymmetry measurement we note that the efficiency for reconstructing 
\Bm\ decays is lower than for \Bzb\ decays by almost a factor of two.
This difference results from a combination of
the different distributions of $X_s$ final states and the different
efficiencies for \KS\ and \Km, \pim\ and \piz.

The efficiency is a rapidly varying function of hadronic mass. 
In the $\Kstar$ region, $M(X_s)<1.1\,\gev$, the efficiency is about 15\,\%, dominated by the high efficiency for 
reconstructing the $\overline{K}^{*0}\to \Km\pip$ mode. For $M(X_s)>1.1\,\gev$ the efficiency decreases
from 5\,\%\ to 1.5\,\%\ as the hadronic mass increases, as shown in Figure~\ref{fig:eff_mass}. 
There are two main reasons for the mass dependence: the multiplicity of the final state 
particles increases with mass, and the angular correlation between them decreases. 
In addition, there are steps in efficiency at $M(X_s)=1.1,2.0$ and $2.4\,\gev$ because
the selection criteria are tightened as the levels of background increase with mass.  

The efficiencies that we obtain from the signal samples are corrected 
for small differences in detection efficiencies between data and Monte Carlo 
events which are determined using control samples. 
The tracking efficiency is reduced by $(0.8 \pm 2.0)\%$ per track, and the \KS\ 
efficiency by $(2 \pm 3)\%$. The photon efficiency is not adjusted, but is assigned 
a 2.5\,\% error per photon. The effect of these adjustments is to reduce 
the overall signal efficiency by $(1.9\pm 6.0)\%$.
The final state distributions are also adjusted by re--weighting the  
signal Monte Carlo events to match the distributions in data (see Section~\ref{sec:missing}). 

Monte Carlo events that are reconstructed in a different final state, 
or with the wrong hadronic mass, are treated as a cross-feed background.
The main sources of cross-feed are events from missing final states which are 
reconstructed in one of the 38 final states, and events from the 38 final states
in which one of the final state particles from the $X_s$ is undetected and 
replaced by a low momentum particle from the other \B.
There is a small contribution from events with multiple candidates in which the 
true candidate is rejected in favour of another candidate with a smaller value of $|\DeltaE|$.
  
The cross-feed background is fitted together with the hadronic \B\ decay backgrounds, 
and allowed to have a component that peaks in the signal region. 
The amount of cross-feed increases as a function of hadronic mass, 
and is largest for the high multiplicity final states. 
As part of our systematic studies we vary the definition of 
cross-feed, transferring events between the 
signal and cross-feed samples. This gives changes in background and 
signal efficiencies which compensate each other within 1\,\%.  

\section{Fitting} \label{sec:fitting}

To extract the signal yield we fit the \mes\ distribution of the data 
with a combination of a Crystal Ball function~\cite{CBfunction} for signal, 
a Novosibirsk function~\cite{Nskfunction} for the peaking backgrounds, 
and ARGUS functions~\cite{ARGUS} for the combinatorial backgrounds. 
The final data fits use \mes\ shapes derived from fits to three sets of 
Monte Carlo samples: signal, continuum background, and \BB\ background 
which includes cross-feed. In all cases we use an unbinned maximum 
likelihood fit.

\begin{figure}[t]
\begin{center}   
\begin{tabular}{c}
  \mbox{\includegraphics[width=3.5in]{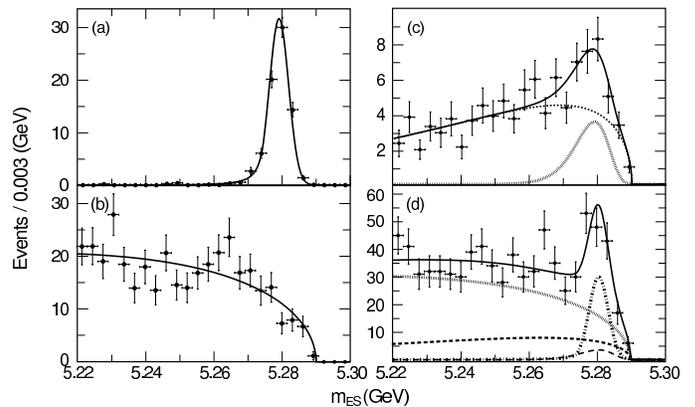}} \\
\end{tabular}
\end{center} 
\vspace{-0.2in}
\caption{Fits to the $M(X_s)=1.4-1.5\,\gev$ bin for (a) correctly reconstructed
signal Monte Carlo events; (b) simulated continuum background;
(c) simulated cross-feed and hadronic $B$ decay backgrounds, where the
contributions from the peaking (dotted) and combinatorial
(dashed) backgrounds are shown separately;
(d) on-peak data, where the contributions from the signal
(dotted-dashed), continuum background(dotted),
peaking \B\ background (long-dashed), and combinatorial \B\ background
(dashed) are shown separately.}
\label{fig:fits} 
\end{figure}

Figure~\ref{fig:fits}(a) shows the fit to the \mes\ distribution of correctly reconstructed 
signal Monte Carlo events with $M(X_s)=1.4-1.5\,\gev$. 
We find no significant variations in the parameters of the Crystal Ball shape 
over the $M(X_s)$ range, so the values used in the
fits to the data are fixed to the weighted average of the results over the full  $M(X_s)$ range:
a width $\sigma = (2.81 \pm 0.05)\,\mev$, a tail parameter $\alpha = 2.17 \pm 0.12$, 
and a slope $n = 0.99 \pm 0.19$. 

Figure~\ref{fig:fits}(b) shows a fit of an ARGUS function to continuum Monte Carlo events in the same bin 
in $M(X_s)$. The amount of continuum background 
increases with $M(X_s)$, and the shape parameter of the 
ARGUS function is a rapidly varying function of $M(X_s)$.  
To cross-check our understanding of the continuum background we also fit
the off-peak data sample. While the overall variations are well reproduced, 
we find small systematic differences in both the shape parameter and the normalization 
which can only be partially accounted for by the difference between the on- and off-peak center-of-mass energies.
We fix the continuum ARGUS shapes and normalizations in the fit to the data to the values
from the continuum Monte Carlo, which vary from bin to bin, and are adjusted for small offsets observed
in comparison to off-peak data. The difference in the
normalization between off-peak data and continuum Monte Carlo is 2.8\,\%.
These offsets are considered to be part of the systematic errors. 

Figure~\ref{fig:fits}(c) shows a fit to the sum of the Monte Carlo predictions for the cross-feed 
and the hadronic \B\ decay backgrounds using an ARGUS function plus a peaking Novosibirsk function. 
In the region $M(X_s)<1.8\,\gev$ the largest contribution comes from the cross-feed, with 
only a small contribution from charmless hadronic $B$ decays. In the region $M(X_s)>1.8\,\gev$ 
the hadronic $\b\to c$ background increases rapidly. 
The peaking background shape is broader than the signal shape, reflecting the less-peaked behavior 
of the backgrounds.
The shape of the Novosibirsk function is determined to have $\sigma = 5.0\,\mev$ and $\tau = -0.295$, 
from a fit to the simulated \BB\ backgrounds over the full $M(X_s)$ range. 

Figure~\ref{fig:fits}(d) shows the fit to the on-peak data in the same mass bin.
In this fit the signal yield and the shape and normalization of the 
combinatorial \BB\ background function are allowed to vary. The signal shape is fixed, 
and the continuum ARGUS function and peaking \BB\ background shapes are fixed. 
The peaking background yield is obtained from the fit shown in  
Figure~\ref{fig:fits}(c).  
Table~\ref{tab:datafits} gives the fitted signal yields, the peaking \BB\ background yields, 
and, as a measure of the goodness of the maximum likelihood fit,
the $\chi^2$ per degree of freedom.

\begin{figure}[htb]
\begin{center}   
   \mbox{\includegraphics[width=3.2in]{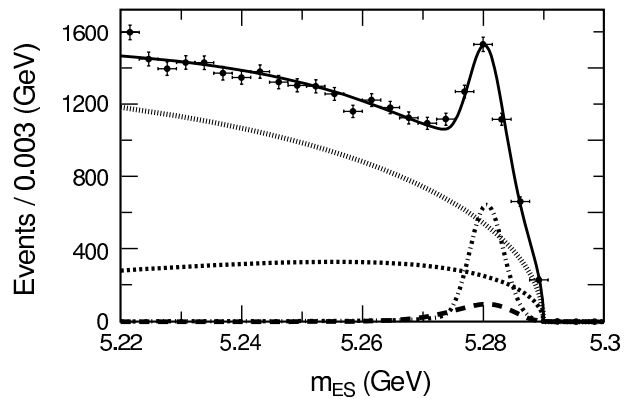}} \\
\end{center} 
\vspace{-0.2in}
\caption{On-peak data fit to the full $M(X_s)$ range, $M(X_s)=0.6-2.8\,\gev$ bin, 
where the contributions from the signal shape (dotted-dashed), fixed continuum ARGUS shape (dotted),
peaking background shape (long-dashed), and combinatorial \BB\ background 
shape (dashed) are shown separately.}
\label{fig:onebin_fit} 
\end{figure}

Figure~\ref{fig:onebin_fit} shows the on--peak data fit to the full $M(X_s)$
range, which gives the yield in the last row of Table~\ref{tab:datafits}.
In this fit the continuum ARGUS shape, and the peaking \BB\ background are taken from fits to 
the full $M(X_s)$ range of the simulated continuum and \BB\ samples. 

The fit procedure has been validated with Monte Carlo studies
to check for biases in the fitting method. 
Systematic errors from the fitting method, including variations in 
the fixed parameters in the fits, are discussed in Section~\ref{sec:syst}.

\begin{table}[htbp]
\caption{Signal yields from the fits to the on-peak data 
and the $\chi^2/$dof from the fits. Also given are the 
peaking background yields from fits to cross-feed and $B\bar{B}$ Monte Carlo.}
\begin{center}
\begin{tabular}{lcD{.}{.}{3.1}cD{.}{.}{2.1}cccD{.}{.}{2.1}cD{.}{.}{1.1}} \hline\hline
\multicolumn{1}{c}{$M(X_s)$} & &  \multicolumn{3}{c}{Data Signal}    & & \multicolumn{1}{c}{Data Fit}       & &  \multicolumn{3}{c}{Peaking Bkg}     \\
\multicolumn{1}{c}{(\gev)}   & &  \multicolumn{3}{c}{yield (events)} & & \multicolumn{1}{c}{$\chi^2/$dof}   & &  \multicolumn{3}{c}{yield (events)} \\ \hline \vspace{-.1in}\\\vspace{.04in}
0.6-0.7  & &   6.5& $\pm$ &7.7     &  & 2.2  &&  0.9 &$\pm$ &3.3      \\\vspace{.04in}
0.7-0.8  & &   5.6& $\pm$ &14.1    &  & 0.8  &&  2.7 &$\pm$ &6.4      \\ \vspace{.04in}
0.8-0.9  & &   416.2& $\pm$& 23.2  &  & 1.5  &&  24.2& $\pm$ &8.5     \\ \vspace{.04in}
0.9-1.0  & &   355.6& $\pm$& 24.9  &  & 0.9  &&  22.7& $\pm$ &10.8    \\ \vspace{.04in}
1.0-1.1  & &   51.3& $\pm$& 19.0   &  & 1.0  &&  14.4& $\pm$ &13.7    \\ \vspace{.04in}
1.1-1.2  & &   33.2& $\pm$& 12.9   &  & 1.2  &&  7.4 &$\pm$ &6.0      \\ \vspace{.04in}
1.2-1.3  & &   83.2& $\pm$ &15.7   &  & 1.1  &&  9.4 &$\pm$ &7.5      \\ \vspace{.04in}
1.3-1.4  & &   101.5& $\pm$ &16.8  &  & 0.8  &&  0.8 &$\pm$ &8.6      \\ \vspace{.04in}
1.4-1.5  & &   72.0& $\pm$& 15.8   &  & 0.8  &&  15.3& $\pm$ &9.1     \\ \vspace{.04in}
1.5-1.6  & &   82.4& $\pm$& 16.5   &  & 1.1  &&  16.1& $\pm$ &11.3    \\ \vspace{.04in}
1.6-1.7  & &   66.1& $\pm$& 16.9   &  & 1.0  &&  5.3 &$\pm$ &11.7     \\\vspace{.04in}
1.7-1.8  & &   54.6& $\pm$& 16.5   &  & 1.3  &&  5.6 &$\pm$ &13.1     \\ \vspace{.04in}
1.8-1.9  & &   76.6& $\pm$& 18.2   &  & 1.1  &&  19.1& $\pm$ &13.7    \\ \vspace{.04in}
1.9-2.0  & &   13.5& $\pm$& 19.5   &  & 1.1  &&  21.3& $\pm$ &14.1    \\ \vspace{.04in}
2.0-2.2  & &   47.5& $\pm$& 21.8   &  & 0.7  &&  19.4& $\pm$ &16.8    \\ \vspace{.04in}
2.2-2.4  & &   52.1& $\pm$& 24.0   &  & 0.7  &&  39.5& $\pm$ &21.0    \\ \vspace{.04in}
2.4-2.6  & &   44.7& $\pm$& 25.6   &  & 0.8  &&  46.8& $\pm$ &20.8    \\ \vspace{.04in}
2.6-2.8  & &   -6.2& $\pm$& 31.9   &  & 1.0  &&  81.0& $\pm$ &26.1    \\
\hline \vspace{-.1in}\\\vspace{.04in}
0.6-2.8  & &  1513.0& $\pm$& 85.1  &  & 1.2  && 464.2& $\pm$ &68.3   \\
\hline \hline
\end{tabular}
\end{center}

\label{tab:datafits}
\end{table} 

\section{$X_s$ Fragmentation and Missing Fractions} \label{sec:missing}

The fragmentation of the $X_s$ system into hadronic final states 
has been modeled using JETSET~\cite{JETSET}. We check this fragmentation by comparing 
the observed data yields in the range $M(X_s)=1.1-2.8\,\gev$ 
with the yields expected from the signal Monte Carlo. 
We do a detailed study by splitting the data and Monte Carlo samples into 10 different 
categories, each containing two to ten of our selected final states. The measured ratios 
of fitted data signal events to reconstructed Monte Carlo signal events
in each category are given in Table~\ref{tab:fragsubset} with their statistical errors. 
\begin{table}[htbp]
\small
\caption{Ratios of data to Monte Carlo yields for various categories of final states. 
These are used to adjust the fragmentation in the signal Monte Carlo.} 
\begin{center}
\begin{tabular}{lc}\hline\hline
Final States & Data/Monte Carlo \\
\hline \vspace{-.1in}\\\vspace{.04in}
$K^-$\pip, \KS~\pim & \er 0.50 0.07 \\\vspace{.04in}
$K^-$\piz, \KS~\piz & \er 0.19 0.12 \\ \hline \vspace{-.1in}\\\vspace{.04in}
$K^-$\pip\pim, \KS~\pip\pim & \er 1.02 0.14 \\\vspace{.04in}
$K^-$\pip\piz, \KS~\pim\piz & \er 1.34 0.24 \\ \hline \vspace{-.1in}\\\vspace{.04in}
$K^-$\pip\pim\pip, \KS~\pip\pim\pim & \er 2.67 0.96 \\\vspace{.04in}
$K^-$\pip\pim\piz, \KS~\pip\pim\piz & \er 1.29 0.61 \\ \hline \vspace{-.1in}\\\vspace{.04in}
$K^-$\piz\piz, \KS~\piz\piz & \\\vspace{.04in}
$K^-$\pip\piz\piz, \KS~\pim\piz\piz & \raisebox{1.5ex}[0pt] {$1.89\pm 1.33$} \\ \hline \vspace{-.1in}\\\vspace{.04in}
$K^-$\pip\pim\pip\pim, \KS~\pip\pim\pip\pim & \\\vspace{.04in}
$K^-$\pip\pim\pip\piz, \KS~\pip\pim\pim\piz & 1.32$^{+1.55}_{-1.32}$ \\\vspace{.04in}
$K^-$\pip\pim\piz\piz, \KS~\pip\pim\piz\piz &  \\ \hline \vspace{-.1in}\\\vspace{.04in}
$K^-\eta$, \KS~$\eta$, $K^-\eta$~\pip & \\\vspace{.04in}
\KS~$\eta$~\pim,  $K^-\eta$~\piz, \KS~$\eta$~\piz & \\\vspace{.04in}
$K^-\eta$~\pip\pim, \KS~$\eta$~\pip\pim & \raisebox{1.5ex}[0pt]{0.83$^{+1.00}_{-0.83}$}  \\ \vspace{.04in}
$K^-\eta$~\pip\piz, \KS~$\eta$~\pim\piz &  \\ \hline \vspace{-.1in}\\\vspace{.04in}
$K^-K^+K^-$, $K^-K^+$\KS & \\\vspace{.04in}
$K^-K^+K^-$\pip, $K^-K^+$\KS~\pim & 0.27$^{+0.54}_{-0.27}$\\\vspace{.04in}
$K^-K^+K^-$\piz, $K^-K^+$\KS~\piz & \\ 
\hline\hline
\end{tabular}
\end{center}

\label{tab:fragsubset}
\end{table}

We note that the rates for the $\kaon\pi$ modes are smaller than the JETSET
prediction. Also interesting is the low ratio for the final states 
with three kaons, where we do not observe a significant signal. These differences 
in fragmentation could be accounted for by changes in the parameters within JETSET, 
and by the addition of resonant contributions, but a detailed study of this requires a larger 
data sample. 

We also make a comparison of the ratio of \KS\ to \Km\ final states in data 
with the same ratio in Monte Carlo. In the range $M(X_s)=1.1-2.8\,\gev$ this gives a double 
ratio of $1.00\pm 0.21$, which is consistent with the assumption of 
isospin symmetry.

The ratios in Table~\ref{tab:fragsubset} are applied as weights to the generated and 
reconstructed signal Monte Carlo events to correct for the observed  
fragmentation. The reduction in the $\kaon\pi$ final states
and the increase in the high multiplicity final states reduces 
the signal efficiencies by 10\,\% to 25\,\%, depending on the $M(X_s)$ bin. 
The weights also lead to an increase in the cross-feed background.   

\begin{figure}[htbp]
\begin{center}
 \begin{tabular}{c}
   \mbox{\includegraphics[width=3.0in]{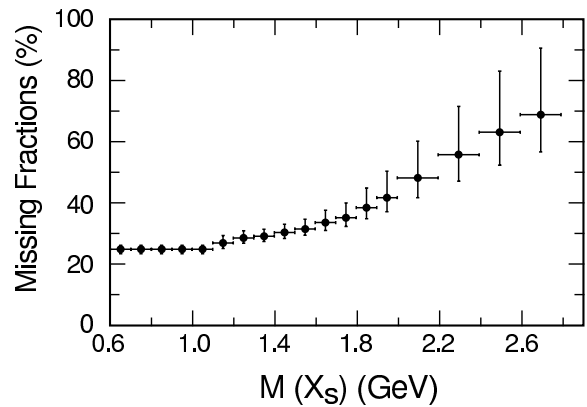}}\\
 \end{tabular}
\end{center}
\caption{Missing fractions with systematic errors as a function of the hadronic mass.}
\label{fig:missing}
\end{figure}

The selected 38 $X_s$ final states do not account for all the states produced in
$\B\to X_s\gamma$. To obtain the total $\B\to X_s\gamma$ branching 
fraction in each $M(X_s)$ bin we need to correct for the fraction of 
missing final states, which increases from 25\,\% to 70\,\% for $M(X_s)=0.6$ to $2.8\,\gev$
(see Fig.~\ref{fig:missing}). 
The 25\,\% fraction missing at all hadronic masses comes from \KL. 
This fraction is equal to the \KS\ sample, with an uncertainty determined by 
our isospin asymmetry measurement (see section X). 

An analysis of the final states generated in the signal Monte Carlo sample 
shows that the largest missing contribution comes from high multiplicity 
final states with a kaon and $\ge 5\pi$. There are also missing contributions
of a few percent in the highest mass bins from higher multiplicity final states 
with $\eta$ mesons, three kaons and baryons. Smaller contributions come from 
rare radiative meson decays and final state radiation.  

We use the results of the fragmentation study to correct for
the missing high multiplicity states with 
$\eta$ and three kaons with the weights found for the observed 
final states. Since the kaon and $4\pi$ modes are consistent with the Monte Carlo 
expectation within a large statistical error, we do not adjust the 
fractions of kaon and $\ge 5\pi$ modes. For these final states, and for missing fractions 
where there is no information from the reconstructed final 
states, we assign systematic errors which allow the missing fraction to 
vary by a factor two relative to the predicted value.
The uncertainty in the missing fractions is the dominant 
systematic error in the high mass bins. 

\section{Systematics} \label{sec:syst}

The following systematic errors are independent of $M(X_s)$. 
There is a 1.1\,\% uncertainty in our knowledge of the number of \B\ mesons in our data 
sample. There is a 2.5\,\% uncertainty in the efficiency of the initial selection of the high 
energy photon, but we apply additional criteria to isolate the high energy photon, as well 
as vetoes if it forms a \piz\ or $\eta$ meson. The efficiency of these additional selections 
is checked using the $\Kstar\gamma$ region as a control sample, and we assign 
an additional uncertainty of 1.5\,\%. The total systematic error independent of $M(X_s)$ 
is 3.1\,\%. 

The systematic errors that depend on $M(X_s)$ are summarized in Table~\ref{tab:summarySyst}.
There are several different categories of systematic errors 
associated with detection efficiencies, fitting, the modeling of peaking backgrounds, 
fragmentation corrections, and the estimation of the missing fractions.

\begin{table}[htbp]
\caption{Contributions to the $M(X_s)$--dependent systematic error on the branching fraction 
from detection efficiency, fitting, peaking background, fragmentation and 
missing fractions, are shown in \%\ as a function of hadronic mass.
The total systematic errors also include a 3.1\,\% systematic error that is 
independent of $M(X_s)$ and not listed in the Table.
\label{tab:summarySyst}}
\begin{center}
\footnotesize
\begin{tabular}{cccD{.}{.}{2.3}D{.}{.}{2.1}cc}\hline\hline
$M(X_s)$  & Detector &  \multicolumn{1}{c}{Fitting}   & \multicolumn{1}{c}{Peaking}  & \multicolumn{1}{c}{Fragmen-} &  \multicolumn{1}{c}{Missing}    & \multicolumn{1}{c}{Total}   \\
(\gev)   & Efficiency &   & \multicolumn{1}{c}{Background}  & \multicolumn{1}{c}{tation} & \multicolumn{1}{c}{Fraction}&    \\ \hline
0.6-0.7 &  5.2  & \multicolumn{1}{c}{$^{+20.8}_{-21.4}$}   &   8.1   &      & \multicolumn{1}{c}{$^{+\hspace{+0.4em}1.8}_{-\hspace{+0.4em}2.0}$}  &   \multicolumn{1}{c}{$^{+23.1}_{-23.6}$}   \\\vspace{.04in}
0.7-0.8 &  5.3  & \multicolumn{1}{c}{$^{+33.1}_{-41.0}$}  &   2.5   &      & \multicolumn{1}{c}{$^{+\hspace{+0.4em}1.8}_{-\hspace{+0.4em}2.0}$}  &   \multicolumn{1}{c}{$^{+33.7}_{-41.4}$}   \\\vspace{.04in}
0.8-0.9 &  5.4  & \multicolumn{1}{c}{$^{+\hspace{+0.4em}2.3}_{-\hspace{+0.4em}2.3}$}   &   1.1   &      & \multicolumn{1}{c}{$^{+\hspace{+0.4em}1.8}_{-\hspace{+0.4em}2.0}$}  &   \multicolumn{1}{c}{$^{+\hspace{+0.4em}6.6}_{-\hspace{+0.4em}6.6}$}    \\\vspace{.04in}
0.9-1.0 &  5.3  & \multicolumn{1}{c}{$^{+\hspace{+0.4em}2.2}_{-\hspace{+0.4em}2.2}$}   &   0.8   &      & \multicolumn{1}{c}{$^{+\hspace{+0.4em}1.8}_{-\hspace{+0.4em}2.0}$}  &   \multicolumn{1}{c}{$^{+\hspace{+0.4em}6.4}_{-\hspace{+0.4em}6.4}$}    \\\vspace{.04in}
1.0-1.1 &  5.3  & \multicolumn{1}{c}{$^{+10.2}_{-10.4}$}  &   3.2   & 13.7 & \multicolumn{1}{c}{$^{+\hspace{+0.4em}2.6}_{-\hspace{+0.4em}2.2}$}  &   \multicolumn{1}{c}{$^{+18.5}_{-18.5}$}    \\\vspace{.04in}
1.1-1.2 &  6.2  & \multicolumn{1}{c}{$^{+11.1}_{-11.2}$}  &   1.7   &  5.4 & \multicolumn{1}{c}{$^{+\hspace{+0.4em}3.4}_{-\hspace{+0.4em}2.5}$}  &   \multicolumn{1}{c}{$^{+14.5}_{-14.4}$}    \\\vspace{.04in}
1.2-1.3 &  6.4  & \multicolumn{1}{c}{$^{+\hspace{+0.4em}5.5}_{-\hspace{+0.4em}5.6}$}   &   1.8   &  4.5 & \multicolumn{1}{c}{$^{+\hspace{+0.4em}3.4}_{-\hspace{+0.4em}2.4}$}  &   \multicolumn{1}{c}{$^{+10.5}_{-10.3}$}    \\\vspace{.04in}
1.3-1.4 &  6.6  & \multicolumn{1}{c}{$^{+\hspace{+0.4em}4.3}_{-\hspace{+0.4em}4.5}$}   &   1.8   &  4.5 & \multicolumn{1}{c}{$^{+\hspace{+0.4em}3.3}_{-\hspace{+0.4em}2.4}$}  &   \multicolumn{1}{c}{$^{+10.0}_{-\hspace{+0.4em}9.8}$}     \\\vspace{.04in}
1.4-1.5 &  6.7  & \multicolumn{1}{c}{$^{+\hspace{+0.4em}4.6}_{-\hspace{+0.4em}4.7}$}   &   2.1   &  5.8 & \multicolumn{1}{c}{$^{+\hspace{+0.4em}3.9}_{-\hspace{+0.4em}2.8}$}  &   \multicolumn{1}{c}{$^{+11.1}_{-10.8}$}    \\\vspace{.04in}
1.5-1.6 &  6.9  & \multicolumn{1}{c}{$^{+\hspace{+0.4em}2.1}_{-\hspace{+0.4em}2.4}$}   &   2.9   &  4.9 & \multicolumn{1}{c}{$^{+\hspace{+0.4em}4.7}_{-\hspace{+0.4em}3.0}$}  &   \multicolumn{1}{c}{$^{+10.5}_{-\hspace{+0.4em}9.9}$}     \\\vspace{.04in}
1.6-1.7 &  7.0  & \multicolumn{1}{c}{$^{+\hspace{+0.4em}3.1}_{-\hspace{+0.4em}3.5}$}   &   4.0   &  5.3 & \multicolumn{1}{c}{$^{+\hspace{+0.4em}6.1}_{-\hspace{+0.4em}3.9}$}  &   \multicolumn{1}{c}{$^{+12.0}_{-11.1}$}    \\\vspace{.04in}
1.7-1.8 &  7.1  & \multicolumn{1}{c}{$^{+\hspace{+0.4em}4.1}_{-\hspace{+0.4em}4.2}$}   &   2.0   &  5.5 & \multicolumn{1}{c}{$^{+\hspace{+0.4em}7.5}_{-\hspace{+0.4em}4.5}$}  &   \multicolumn{1}{c}{$^{+12.7}_{-11.2}$}    \\\vspace{.04in}
1.8-1.9 &  7.2  & \multicolumn{1}{c}{$^{+\hspace{+0.4em}3.2}_{-\hspace{+0.4em}3.4}$}   &   2.2   &  6.5 & \multicolumn{1}{c}{$^{+10.6}_{-\hspace{+0.4em}5.9}$} &   \multicolumn{1}{c}{$^{+15.0}_{-12.2}$}    \\\vspace{.04in}
1.9-2.0 &  7.1  & \multicolumn{1}{c}{$^{+32.6}_{-32.4}$}  &   9.3   &  5.4 & \multicolumn{1}{c}{$^{+15.0}_{-\hspace{+0.4em}8.0}$} &   \multicolumn{1}{c}{$^{+38.2}_{-35.8}$}    \\\vspace{.04in}
2.0-2.2 &  7.2  & \multicolumn{1}{c}{$^{+11.6}_{-\hspace{+0.4em}9.4}$}   &   5.1   &  6.2 & \multicolumn{1}{c}{$^{+23.4}_{-12.7}$}&   \multicolumn{1}{c}{$^{+28.3}_{-19.2}$}    \\\vspace{.04in}
2.2-2.4 &  7.5  & \multicolumn{1}{c}{$^{+10.8}_{-\hspace{+0.4em}9.1}$}   &   8.9   &  7.1 & \multicolumn{1}{c}{$^{+36.2}_{-19.8}$}&   \multicolumn{1}{c}{$^{+40.2}_{-25.8}$}    \\\vspace{.04in}
2.4-2.6 &  7.5  & \multicolumn{1}{c}{$^{+\hspace{+0.4em}7.6}_{-\hspace{+0.4em}7.6}$}   &   11.1  &  9.3 & \multicolumn{1}{c}{$^{+55.2}_{-29.7}$}&   \multicolumn{1}{c}{$^{+58.1}_{-34.8}$}    \\\vspace{.04in}
2.6-2.8 &  7.9  & \multicolumn{1}{c}{$^{+68.3}_{-96.1}$}  &   65.7  & 11.7 & \multicolumn{1}{c}{$^{+71.3}_{-39.9}$}&   \multicolumn{1}{c}{$^{+119.0}_{-124.0}$}   \\
\hline \vspace{-.1in}\\\vspace{.04in}	 
0.6-2.8 &  6.1  & \multicolumn{1}{c}{$^{+\hspace{+0.4em}3.2}_{-\hspace{+0.4em}3.7}$}   &   1.6   &  5.9 & \multicolumn{1}{c}{$^{+13.8}_{-\hspace{+0.4em}7.6}$} & \multicolumn{1}{c}{$^{+16.7}_{-12.2}$}    
\\
\hline\hline
\end{tabular}
\end{center}
\end{table}

The detection efficiency errors come from our knowledge of the 
differences between Monte Carlo and data obtained from control samples. 
They reflect the accuracy of our modeling of the detector response, and include  
tracking efficiency (2\,\%), particle identification for charged kaons (1\,\%), 
\KS\ reconstruction (3\,\%), and photon detection efficiency (2.5\,\%). 
These differences give not only systematic errors but also shifts in the signal 
detection efficiencies, as mentioned in Section~\ref{sec:efficiency}.
The fitting errors come from varying the fixed parameters in the data fits.
The signal peak position in $m_{ES}$ is varied by $\pm\,0.4\,\mev$, and the width 
by $\pm\,0.1\,\mev$ to allow for a possible variation as a function of $M(X_s)$. 
The width is varied in a correlated fashion with the $\alpha$ and tail parameters 
of the Crystal Ball function. 
These changes in the signal shape alter the signal yields by between 0 and 5 events 
in each $M(X_s)$ bin. The peaking background shape is varied  
within a range allowed by the cross-feed and \BB\ Monte Carlo samples. 
This gives smaller changes in the signal yield of about one event in each bin.
The continuum shape and normalization are varied by the difference between 
the continuum Monte Carlo and the off-peak data, but this gives very small changes 
in the signal yields. 

The peaking background normalization error is treated separately.
In the region $M(X_s)<1.8\,\gev$ the main uncertainty comes  
from the charmless hadronic decays $\B\to X_s\piz$. While some of these have been measured, 
others are yet to be observed. Combining the effects of the measured and the unmeasured modes 
we assign an uncertainty of 50\,\% to the normalization of this contribution in the low $M(X_s)$ region.
In the region $M(X_s)>1.8\,\gev$ the background is primarily from
hadronic $\b\to c$ decays, such as $\B\to D^*\rho^-$. By taking the weighted average
of the uncertainties on the branching fractions of the components of this background, 
we assign an uncertainty of 15\,\%\ to the normalization of this contribution in
the high $M(X_s)$ region.
The cross-feed background depends on the modeling of the $\B\to X_s\gamma$ process. 
We use the fragmentation weights and the measured spectral shape to adjust the 
signal Monte Carlo to match the data, correct our prediction for the 
peaking cross-feed background, and assign an uncertainty to this correction.

As discussed in Section~\ref{sec:missing}, we have studied the differences in
fragmentation between data and signal Monte Carlo, and re-weighted the signal Monte Carlo to correct
for the differences found. The weighting factors are listed by category
in Table~\ref{tab:fragsubset}. We vary the weights by their
statistical errors to evaluate the fragmentation systematics.   
For the $\eta$ and $\kaon+4\pi$ categories the weights are consistent with one 
(with larger errors), and we restrict the range to be between $0.5$ and $2.0$.
Each weight is varied separately, and assumed to be uncorrelated 
with the other weights. The fragmentation errors are limited by the statistics
of the data sample.
In the bin $M(X_s)=1.0-1.1\,\gev$ the efficiency is computed from the average 
of the resonant and non-resonant model efficiencies, 
and we take the difference between them as 
an additional fragmentation systematic of 12.8\,\% in this bin.

The estimation of the missing fractions is the largest systematic error. 
It is determined by varying the missing fractions within the ranges shown in Figure~\ref{fig:missing}.  
We vary all the uncertainties in the missing fractions together, 
either increasing them all, or decreasing them all. 
When we do this we adjust the sum of the reconstructed final states so that the 
total $\B\to X_s\gamma$ rate is unchanged, i.e. we actually adjust the relative proportions 
of reconstructed and missing final states.

\section{Branching Fraction Results} \label{sec:bf}
The branching fractions in each hadronic mass bin are obtained using 
the signal efficiencies shown in Figure~\ref{fig:eff_mass}, the signal yields given in Table~\ref{tab:datafits}, 
the fraction of reconstructed final states (which is 1 minus the fraction of missing 
final states shown in Figure~\ref{fig:missing}), 
and the total number of \BB\ pairs in the sample. 
The systematic studies which affect each of these quantities were discussed in Section~\ref{sec:syst}.  

\begin{figure}[htbp]
\begin{center}
 \begin{tabular}{c}
   \mbox{\includegraphics[width=3in]{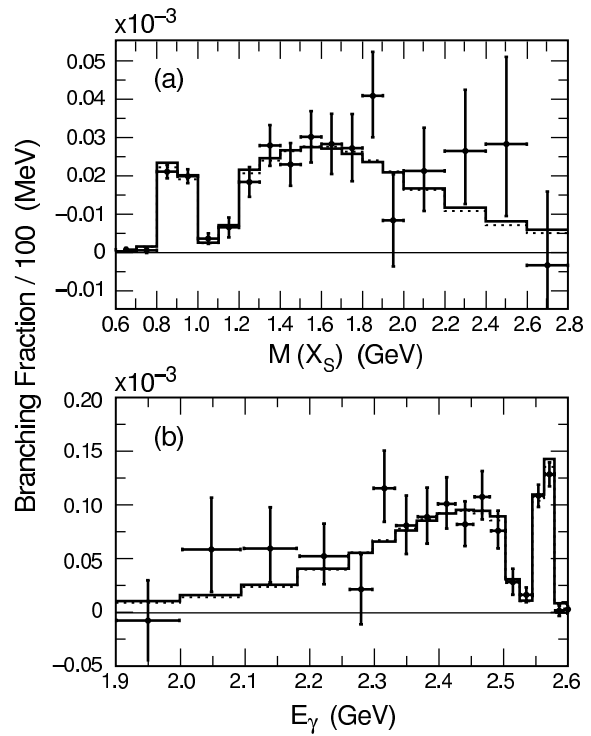}}\\
 \end{tabular}
\end{center}
\caption{The hadronic mass spectrum (a), and the
photon energy spectrum (b). The data points are compared to theoretical predictions (histograms) obtained 
using the shape function (solid line) and
kinetic (dashed line) schemes.}
\label{fig:spectrum}
\end{figure}

Table~\ref{tab:results} shows the results for the branching fraction in each hadronic 
mass bin, as well as the result for the whole mass range.
Table~\ref{tab:resultsgamma} shows the corresponding branching fractions in bins of the 
photon energy using Equation (\ref{eq:egamma})
to translate between hadronic mass and photon energy. We have 
taken into account the different bin sizes in transforming between ${\cal B}(M(X_s))$ and 
${\cal B}(E_\gamma)$.
The corresponding hadronic mass and photon energy spectra are shown in Figure~\ref{fig:spectrum},
where theoretical predictions are shown which will be discussed in Section~\ref{sec:spectrum}.
The hadronic mass resolution of $10\,\mev$ converts into an $E_{\gamma}$ resolution
of 1-5\,\mev, and the $\Kstar$ peak can be clearly seen in both 
the hadronic mass and photon energy spectra.

\begin{table}[htbp]
\caption{Branching fractions in bins of hadronic mass with statistical and systematic errors.
The bottom line shows the total branching fraction obtained from the separate fit to the data 
over the full $M(X_s)$ range, and not from the sum of the individual bins.}
\begin{center}
\begin{tabular}{c D{.}{.}{7.2}cD{.}{.}{2.0}c}
\hline\hline
$M(X_s)$~  (\gev)~~~~ & \multicolumn{4}{c}{  ${\cal B}(M(X_s))$/100\mev  ~~$(10^{-6})$ } \\ \hline \vspace{-.1in}\\\vspace{.04in}
0.6 - 0.7 &  0.4 &$\pm$&  0.5 & \multicolumn{1}{c}{$^{+\hspace{+0.4em}0.1}_{-\hspace{+0.4em}0.1}$}   \\\vspace{.04in}
0.7 - 0.8 &  0.3 &$\pm$&  0.7 & \multicolumn{1}{c}{$^{+\hspace{+0.4em}0.1}_{-\hspace{+0.4em}0.1}$}   \\\vspace{.04in}
0.8 - 0.9 &  20.8 &$\pm$&  1.2  & \multicolumn{1}{c}{$^{+\hspace{+0.4em}1.3}_{-\hspace{+0.4em}1.3}$}     \\\vspace{.04in}
0.9 - 1.0 &  19.6 &$\pm$&  1.4  & \multicolumn{1}{c}{$^{+\hspace{+0.4em}1.2}_{-\hspace{+0.4em}1.2}$}     \\\vspace{.04in}
1.0 - 1.1 &  3.3  &$\pm$&  1.2  & \multicolumn{1}{c}{$^{+\hspace{+0.4em}0.6}_{-\hspace{+0.4em}0.6}$}     \\\vspace{.04in}
1.1 - 1.2 &  6.2  &$\pm$&  2.4  & \multicolumn{1}{c}{$^{+\hspace{+0.4em}0.9}_{-\hspace{+0.4em}0.9}$}     \\\vspace{.04in}
1.2 - 1.3 &  18.1 &$\pm$&  3.4  & \multicolumn{1}{c}{$^{+\hspace{+0.4em}1.9}_{-\hspace{+0.4em}1.9}$}     \\\vspace{.04in}
1.3 - 1.4 &  27.6 &$\pm$&  4.6  & \multicolumn{1}{c}{$^{+\hspace{+0.4em}2.8}_{-\hspace{+0.4em}2.7}$}     \\\vspace{.04in}
1.4 - 1.5 &  22.6 &$\pm$&  5.0  & \multicolumn{1}{c}{$^{+\hspace{+0.4em}2.5}_{-\hspace{+0.4em}2.5}$}     \\\vspace{.04in}
1.5 - 1.6 &  29.8 &$\pm$&  6.0  & \multicolumn{1}{c}{$^{+\hspace{+0.4em}3.1}_{-\hspace{+0.4em}3.0}$}     \\\vspace{.04in}
1.6 - 1.7 &  28.0 &$\pm$&  7.2  & \multicolumn{1}{c}{$^{+\hspace{+0.4em}3.3}_{-\hspace{+0.4em}3.1}$}     \\\vspace{.04in}
1.7 - 1.8 &  26.9 &$\pm$&  8.1  & \multicolumn{1}{c}{$^{+\hspace{+0.4em}3.4}_{-\hspace{+0.4em}3.0}$}     \\\vspace{.04in}
1.8 - 1.9 &  40.6 &$\pm$&  9.7  & \multicolumn{1}{c}{$^{+\hspace{+0.4em}6.1}_{-\hspace{+0.4em}5.0}$}     \\\vspace{.04in}
1.9 - 2.0 &   8.0 &$\pm$&  11.7 & \multicolumn{1}{c}{$^{+\hspace{+0.4em}3.1}_{-\hspace{+0.4em}2.9}$}     \\\vspace{.04in}
2.0 - 2.2 &  21.0 &$\pm$&  9.6  & \multicolumn{1}{c}{$^{+\hspace{+0.4em}5.9}_{-\hspace{+0.4em}4.0}$}     \\\vspace{.04in}
2.2 - 2.4 &  26.1 &$\pm$&  12.0 & \multicolumn{1}{c}{$^{+10.5}_{-\hspace{+0.4em}6.7}$}    \\\vspace{.04in}
2.4 - 2.6 &  28.0 &$\pm$&  16.0 & \multicolumn{1}{c}{$^{+16.2}_{-\hspace{+0.4em}9.7}$}    \\\vspace{.04in}
2.6 - 2.8 &  -3.7 &$\pm$&  18.8 & \multicolumn{1}{c}{$^{+\hspace{+0.4em}4.4}_{-\hspace{+0.4em}4.5}$}     \\ 
\hline\hline \vspace{-.1in}\\\vspace{.04in}
&  \multicolumn{4}{c}{ ${\cal B}~(10^{-6})$}  \\ \hline \vspace{-.1in}\\\vspace{.04in}
0.6 - 2.8 &  327.0  &$\pm$&  18.0   & \multicolumn{1}{c}{$^{+55.0}_{-40.0}$}    \\
\hline\hline
\end{tabular}
\end{center}
\label{tab:results}
\end{table}

\begin{table}[htbp]
\caption{Branching fractions in bins of photon energy with statistical and systematic errors.}
\begin{center}
\begin{tabular}{c D{.}{.}{7.1}cD{.}{.}{1.0}c}
\hline\hline
~$E_\gamma$~~(\gev)~~~ &  \multicolumn{4}{c}{ ${\cal B}(E_{\gamma})$/100\mev ~~($10^{-6})$} \\ \hline \vspace{-.1in}\\\vspace{.04in}
2.593 - 2.606 &  3.3  &$\pm$&  4.0   & \multicolumn{1}{c}{$^{+\hspace{+0.4em}0.8}_{-\hspace{+0.4em}0.8}$}  \\\vspace{.04in}
2.579 - 2.593 &  1.9  &$\pm$& 4.9    & \multicolumn{1}{c}{$^{+\hspace{+0.4em}0.6}_{-\hspace{+0.4em}0.8}$}   \\\vspace{.04in}
2.563 - 2.579 &  129.2  &$\pm$& 7.2  & \multicolumn{1}{c}{$^{+\hspace{+0.4em}8.1}_{-\hspace{+0.4em}8.1}$}   \\\vspace{.04in}
2.545 - 2.563 &  108.9  &$\pm$& 7.6  & \multicolumn{1}{c}{$^{+\hspace{+0.4em}6.6}_{-\hspace{+0.4em}6.6}$}   \\\vspace{.04in}
2.525 - 2.545 &  16.7 &$\pm$& 6.2    & \multicolumn{1}{c}{$^{+\hspace{+0.4em}3.2}_{-\hspace{+0.4em}3.2}$}   \\\vspace{.04in}
2.503 - 2.525 &  28.6 &$\pm$& 11.1   & \multicolumn{1}{c}{$^{+\hspace{+0.4em}4.1}_{-\hspace{+0.4em}4.1}$}   \\\vspace{.04in}
2.480 - 2.503 &  76.3 &$\pm$& 14.4   & \multicolumn{1}{c}{$^{+\hspace{+0.4em}8.0}_{-\hspace{+0.4em}7.8}$}  \\\vspace{.04in}
2.454 - 2.480 &  107.8  &$\pm$& 17.9 & \multicolumn{1}{c}{$^{+10.8}_{-10.6}$} \\\vspace{.04in}
2.427 - 2.454 &  82.4 &$\pm$& 18.1   & \multicolumn{1}{c}{$^{+\hspace{+0.4em}9.2}_{-\hspace{+0.4em}8.9}$}  \\\vspace{.04in}
2.397 - 2.427 &  101.6  &$\pm$& 20.3 & \multicolumn{1}{c}{$^{+10.6}_{-10.1}$} \\\vspace{.04in}
2.366 - 2.397 &  89.5 &$\pm$& 22.9   & \multicolumn{1}{c}{$^{+10.7}_{-\hspace{+0.4em}9.9}$}  \\\vspace{.04in}
2.333 - 2.366 &  81.3 &$\pm$& 24.6   & \multicolumn{1}{c}{$^{+10.3}_{-\hspace{+0.4em}9.1}$}  \\\vspace{.04in}
2.298 - 2.333 &  115.8  &$\pm$& 27.6 & \multicolumn{1}{c}{$^{+17.4}_{-14.1}$} \\\vspace{.04in}
2.261 - 2.298 &  21.8 &$\pm$& 31.6   & \multicolumn{1}{c}{$^{+\hspace{+0.4em}8.3}_{-\hspace{+0.4em}7.8}$}   \\\vspace{.04in}
2.181 - 2.261 &  52.7 &$\pm$& 24.2   & \multicolumn{1}{c}{$^{+14.9}_{-10.1}$}  \\\vspace{.04in}
2.094 - 2.181 &  60.0 &$\pm$& 27.6   & \multicolumn{1}{c}{$^{+24.1}_{-15.5}$} \\\vspace{.04in}
1.999 - 2.094 &  59.0 &$\pm$& 33.8   & \multicolumn{1}{c}{$^{+34.3}_{-20.5}$} \\\vspace{.04in}
1.897 - 1.999 &  -7.1 &$\pm$& 36.7   & \multicolumn{1}{c}{$^{+\hspace{+0.4em}8.5}_{-\hspace{+0.4em}8.8}$}  \\
\hline\hline
\end{tabular}
\end{center}
\label{tab:resultsgamma}
\end{table}

\section{Isospin Asymmetry} \label{sec:isospin}

We define the isospin asymmetry as the ratio:
\begin{equation}
\label{eq:isospin}
	\Delta_{0- } 
	= {{{\Gamma}(\Bzb \to X_{s\bar{d}}\gamma) - {\Gamma}(\Bm\to X_{s\bar{u}}\gamma)}
	\over{{\Gamma}(\Bzb \to X_{s\bar{d}}\gamma) + {\Gamma}(\Bm\to X_{s\bar{u}}\gamma)}}.
\end{equation}
The Standard Model predicts no isospin symmetry-breaking from the dominant penguin
diagram for $\B \to X_s \gamma$. Isospin symmetry-breaking effects occur at 
order $\Lambda/m_{\b}$ in the heavy quark expansion \cite{IsospinKN}, 
due to annihilation contributions from four-quark operators, 
the chromo-magnetic dipole operator and charm penguins. 
For the exclusive decays $\B\to K^*\gamma$, the Standard Model predicts 
a positive value of $\Delta_{0-}$ between 5 and 10\,\%~\cite{IsospinKN},  
but new physics beyond the Standard Model could enhance the isospin 
breaking effects. Measurements of the $\B\to K^*\gamma$ isospin asymmetry from 
\babar\ and BELLE are consistent with the predictions of the Standard Model \cite{Kstar}. 

We split the 38 modes into charged and neutral \B\ decays, re-fit the data, 
and calculate the separate efficiencies and total branching fractions.
While the signal detection efficiencies are almost a factor of two lower for the 
$B^-$ decays, the backgrounds and missing fractions are symmetric. 
Comparing the charged and neutral branching fraction measurements, using
the lifetime ratio, $\tau(\Bm)/\tau(\Bz)=1.086\pm 0.017$~\cite{PDG}, and 
our recent measurement of the production ratio of charged and neutral \B\ events 
at the \FourS, $\Bzb\ /\Bm= 1.006 \pm 0.048$~\cite{ProdRatio}, 
gives the isospin asymmetry over the range $M(X_s)=0.6-2.8\,\gev$:
\begin{displaymath}
\Delta_{0-} = -0.006 \pm 0.058 \pm 0.009 \pm 0.024
\end{displaymath} 

The errors are statistical, systematic and due to the production ratio $\Bzb\ /\Bm$,
respectively.
Most of the systematic errors on the branching fractions cancel in the ratio.
The residual systematic errors that are relevant to the isospin asymmetry  
are $\pm 0.001$ from the detection efficiency corrections, and $+0.004$ from a 
study of the fragmentation differences between charged and neutral \B\ modes
in which we allow for different \Bm\ and \Bzb\ weights. The
largest contribution to our systematic error is
$\pm 0.008$ due to the uncertainty in the lifetime ratio.

\section{Fits to the Spectrum, Extraction of $m_\b$ and $\mu_{\pi}^2$, and Inclusive Branching
Fraction} 
\label{sec:spectrum}
In the following section, we evaluate the results in the context of recent QCD calculations 
in the shape function~\cite{TheoryBF3, ShapeFunction4, ShapeFunction5, ShapeFunction6, N1, N2} and kinetic~\cite{BBU} schemes.
From a fit to the spectrum we evaluate the \b--quark
mass $m_\b(\mu) $, and the kinetic--energy parameter $\mu_\pi^2(\mu)$, 
where the reference scales are taken to be $\mu=1.5\,\gev$ in the shape function scheme,
and $\mu=1.0\,\gev$ in the kinetic scheme. We have set the chromomagnetic operator $\mu_G = 0.35\gev ^2 $,
and the Darwin and spin--orbit operators to $\rho_D=0.2\gev^3$ and $\rho_{LS}=-0.09\gev^3$
in the kinetic scheme~\cite{Uraltsev}.
The photon and hadronic mass spectra are equivalent, so we can fit either one to extract the 
heavy--quark parameters.

We use a $\chi^2$ method to fit the spectrum and find the best values of the parameters $m_\b$ and 
$\mu_{\pi}^2$, adding an additional constraint on the normalization of the spectrum 
from the value of ${\cal B}(\b\to s\gamma)$ measured over the full range $M(X_s)=0.6-2.8\,\gev$.
The fit method takes into account the asymmetry of the systematic errors and the large 
bin-to-bin correlations.

The spectrum is fit using the expected spectral shape from the two schemes, 
except that at low hadronic mass we replace the inclusive theoretical distribution with a
Breit--Wigner to represent the $\Kstar$ resonance. The transition point
between the $\Kstar$ and inclusive distributions 
is a free parameter of the fit. As a cross--check we have also performed fits to the 
spectrum where we treat the $\Kstar$ region $M(X_s)=0.6-1.2\,\gev$ as a single bin. 
In this case a Breit-Wigner shape and a transition point are unnecessary to describe the data.
The results of this cross--check agree with the default fits with the $\Kstar$ included.

The signal Monte Carlo, which is used to determine the signal efficiencies and the 
cross-feed background, depends on the mass scheme used and the corresponding 
heavy--quark parameters. Starting from the initial fits to the 
measured spectrum we modify the signal Monte Carlo with the fitted parameters, and revise our 
estimates of efficiencies and cross--feed. We re--fit the data, re--calculate 
the branching fractions and then re--fit the spectrum. This procedure leads to small changes 
in the heavy--quark parameters which are well within the errors. 

The final results after the re-fitting are shown in Table~\ref{tab:spectrumfit} 
for $m_\b$ and $\mu_\pi^2$. 
In the shape function scheme we present results using three different models for the shape function~\cite{ShapeFunction6}. 
The exponential and hyperbolic models give very similar results, and the exponential model is taken to be the default  
in this scheme. Slightly different results are obtained with the Gaussian model.
In the kinetic scheme, we use the two functions provided by the authors, and quote the average~\cite{Uraltsev}.  
We show the fits to the measured hadronic mass and photon energy spectrum in Figure~\ref{fig:spectrum}, from which it can be seen 
that the spectrum is well described and the difference between the two schemes is small.
The central values and error ellipses for the shape function scheme of the fitted heavy--quark parameters are 
shown in Figure~\ref{fig:ellipses} (the points of the exponential model ellipse are given in
Appendix~\ref{sec:appendix}).
For comparison with previous measurements we have also fitted the spectrum to 
the older model of Kagan and Neubert~\cite{KaganNeubert}. 
We find $m_\b = (4.79^{+0.06}_{-0.10})\gev$ for the \b quark mass and 
$\lambda_1 = (-0.24^{+0.09}_{-0.18})\gev^2$ for the kinetic parameter, with a 94\,\%\
correlation between them.

\begin{table}[htbp]
\caption{Heavy--quark parameters $m_\b$ and $\mu_\pi^2$ from fits to the spectrum using 
the three different models of the shape function scheme  
and using the kinetic scheme. For each scheme the first two columns are from a fit 
with just statistical errors, and the last two columns from a fit with statistical 
and systematic errors.
The correlation coefficients between the two parameters are -94\% and -92\% 
in the two schemes.}
\begin{center}
\begin{tabular}{l c c c c}
\hline\hline
Theoretical  & \multicolumn{2}{c}{Stat. Errors} & \multicolumn{2}{c}{Stat.+Syst. Errors} \\
Scheme   & $m_\b (\gev)$ & $\mu_\pi^2 (\gev^2)$  & $m_\b (\gev)$ & $\mu_\pi^2 (\gev^2)$ \\ \hline \vspace{-.1in}\\\vspace{.04in}
%&  &  & \\ 
Shape Function &  &  & \\ \vspace{.04in}
Exponential  & $4.65\pm 0.04$ & $0.19\pm 0.06$ & $4.67\pm 0.07$ & $0.16^{+0.10}_{-0.08}$  \\ \vspace{.04in}
Hyperbolic   & $4.64\pm 0.04$ & $0.20\pm 0.06$ & $4.67\pm 0.07$ & $0.17^{+0.10}_{-0.09}$  \\ \vspace{.04in}
Gaussian     & $4.68\pm 0.04$ & $0.12\pm 0.06$ & $4.73^{\hspace{+0.4em}+\hspace{+0.4em}0.06}_{\hspace{+0.4em}-\hspace{+0.4em}0.07}$ & $0.07^{+0.09}_{-0.06}$  \\ 
%&  &  & \\ 
\hline \vspace{-.1in}\\\vspace{.04in}
%&  &  & \\
Kinetic & $4.67\pm 0.04$ & $0.32^{\hspace{+0.4em}+\hspace{+0.4em}0.07}_{\hspace{+0.4em}-\hspace{+0.4em}0.04}$ & $4.70^{\hspace{+0.4em}+\hspace{+0.4em}0.04}_{\hspace{+0.4em}-\hspace{+0.4em}0.08}$ & $0.29^{+0.09}_{-0.04}$ \\
%&  &  & \\
\hline\hline
\end{tabular}
\end{center}
\label{tab:spectrumfit}
\end{table}

\begin{figure}[htbp]
\begin{center}
  \mbox{\includegraphics[width=3.0in]{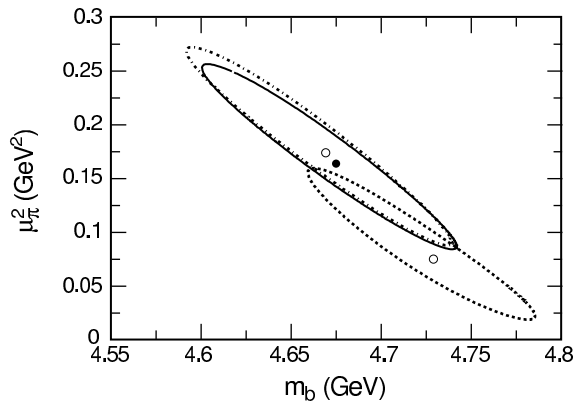}}\\
\end{center}
\caption{Error ellipses corresponding to $\Delta\chi^2 = 1$ from the fit to the 
spectrum in the shape function scheme using the exponential
(solid line), hyperbolic (dotted-dashed) and Gaussian (dashed line) models of the shape function.
Both statistical and systematic errors have been taken into account.}
\label{fig:ellipses}
\end{figure}

The inclusive branching fraction is obtained from a fit to the \mes\
distribution of data events over the full $M(X_s)$ range, corresponding to $E_{\gamma}>1.90\gev$
(Figure~\ref{fig:onebin_fit}).
In the shape function scheme we obtain  
${\cal B}(\b\to s\gamma) = (3.27\pm 0.18 ^{+0.55 +0.04}_{-0.40 -0.06})\times 10^{-4}$
and in the kinetic scheme
${\cal B}(\b\to s\gamma) = (3.27\pm 0.18 ^{+0.55 +0.04}_{-0.40 -0.12})\times 10^{-4}$.
The errors are respectively statistical, systematic and due to the 
variation of the shape parameters.
We quote the average of the results from the two theoretical schemes:
\begin{displaymath}
{\cal B}(\b\to s\gamma) = (3.27\pm 0.18 ^{+0.55 +0.04}_{-0.40 -0.09})\times 10^{-4}, \hspace{0.2cm}
E_{\gamma}>1.9~\gev
\end{displaymath}
The branching fraction can be extrapolated to a lower photon energy using 
the fits to the spectrum. Again we quote the average of the two schemes:
\begin{displaymath}
{\cal B}(\b\to s\gamma) = (3.35\pm 0.19 ^{+0.56 +0.04}_{-0.41 -0.09})\times 10^{-4}, \hspace{0.2cm}
E_{\gamma}>1.6~\gev
\end{displaymath}
where the small uncertainties from the extrapolations in the two schemes 
enter into the model error through the variation of the fitted parameters.
 
\section{Moments of the photon energy spectrum} 
\label{sec:moments}
The first moment is defined as the average of the photon energy spectrum, $\langle E_\gamma\rangle$, while higher moments
are defined as $\langle(E_\gamma-\langle E_\gamma\rangle)^N\rangle$, where $N$ is the order of the moment under investigation.
The values of the moments depend on the range of the photon energy spectrum used to calculate them.
We vary the range by considering five different minimum photon energies
$E_\gamma^{min}=1.897$, 1.999, 2.094, 2.181 and 2.261\,\gev. These values correspond to the boundaries of the
highest bins in hadronic mass, $M(X_s)=2.8$, 2.6, 2.4, 2.2 and 2.0\,\gev. 

\begin{figure}[htbp]
\begin{center}
\begin{tabular}{c}
  \mbox{\includegraphics[width=3.0in]{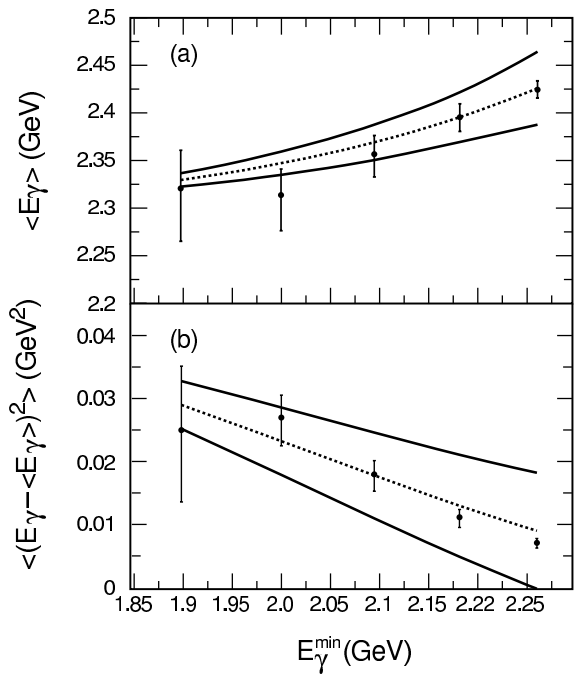}}\\
 \end{tabular}
\end{center}
\caption{First (a), and second (b) moments as a function of the minimum photon energy.
The dotted lines show the predicted central values based on fits to 
the $b\to c\ell\nu$ moments~\cite{Vcb}, and the solid lines the theoretical 
uncertainties from the kinetic scheme~\cite{BBU, Uraltsev}.}
\label{fig:moments}
\end{figure}

The results for the first, second and third moments of our photon energy spectrum
as a function of the minimum photon energy are shown in Table~\ref{tab:moments}.
The third moments can be 
used to test the predictions for the $\B \to X_s \gamma$ decay spectrum
by dressed gluon exponentiation~\cite{DGE}, but for now they are statistically
limited.
Figure~\ref{fig:moments} shows the first and second moments as a function of the minimum photon energy,
together with the predictions from Ref.~\cite{BBU}, which uses the parameters $m_b=4.61\,\gev$ and 
$\mu_{\pi}^2=0.45\,\gev^2$ that are obtained from fits to the 
$b\to c\ell\nu$ moments~\cite{Vcb}. The solid lines represent the band allowed by
theoretical uncertainties~\cite{Uraltsev}. 
While the experimental errors decrease rapidly as the minimum photon energy
is raised, the theoretical errors increase due to the ``bias'' corrections described 
in Ref.~\cite{BBU}. The agreement between the $b\to s\gamma$ and $b\to c\ell\nu$ moments is 
good, and well within the expected theoretical uncertainties. 
This demonstrates a non-trivial consistency between two different classes
of inclusive $b$ decays.   

\begin{table}[htbp]
\caption{\label{tab:moments} {First, second and third moments of the photon energy spectrum as a function of the
minimum photon energy with statistical and systematic errors.}}
\begin{center}
\begin{tabular}{c c  }
\hline\hline \vspace{-.1in}\\\vspace{.04in}
$E_\gamma^{min}$ (\gev) & $\langle E_\gamma\rangle$ (\gev)  \\ \hline \vspace{-.1in}\\\vspace{.04in}
1.897  & $ 2.321  \pm ~  0.038 ~^{+0.017}_{-0.038}$    \\\vspace{.04in}
1.999  & $ 2.314  \pm ~  0.023 ~^{+0.014}_{-0.029}$    \\\vspace{.04in}
2.094  & $ 2.357  \pm ~  0.017 ~^{+0.007}_{-0.017}$    \\\vspace{.04in}
2.181  & $ 2.396  \pm ~  0.013 ~^{+0.003}_{-0.009}$    \\\vspace{.04in}
2.261  & $ 2.425  \pm ~  0.009 ~^{+0.002}_{-0.004}$    \\
\hline \hline \vspace{-.1in}\\\vspace{.04in}
$E_\gamma^{min}$ (\gev) & $\langle(E_\gamma-\langle E_\gamma\rangle)^2\rangle$ ($\gev^2$)  \\ \hline \vspace{-.1in}\\\vspace{.04in}
1.897 & $ 0.0253  \pm ~  0.0101 ~  ^{+0.0041}_{-0.0028}$   \\\vspace{.04in}
1.999 & $ 0.0273  \pm ~  0.0037 ~  ^{+0.0015}_{-0.0015}$   \\\vspace{.04in}
2.094 & $ 0.0183  \pm ~  0.0023 ~  ^{+0.0010}_{-0.0007}$   \\\vspace{.04in}
2.181 & $ 0.0115  \pm ~  0.0014 ~  ^{+0.0005}_{-0.0003}$   \\\vspace{.04in}
2.261 & $ 0.0075  \pm ~  0.0007 ~  ^{+0.0002}_{-0.0002}$   \\
\hline \hline \vspace{-.1in}\\\vspace{.04in}
$E_\gamma^{min}$ (\gev) & $\langle(E_\gamma-\langle E_\gamma\rangle)^3\rangle$ ($\gev^3$)  \\ \hline \vspace{-.1in}\\\vspace{.04in}
1.897 & $ -0.0006  \pm ~  0.0012 ~  ^{+0.0009}_{-0.0002}$     \\\vspace{.04in}
1.999 & $ -0.0009  \pm ~  0.0006 ~  ^{+0.0010}_{-0.0004}$     \\\vspace{.04in}
2.094 & $ -0.0005  \pm ~  0.0003 ~  ^{+0.0004}_{-0.0001}$     \\\vspace{.04in}
2.181 & $ -0.0001  \pm ~  0.0001 ~  ^{+0.0001}_{-0.0000}$     \\\vspace{.04in}
2.261 & $ +0.0001  \pm ~  0.0001 ~  ^{+0.0000}_{-0.0000}$    \\
\hline\hline
\end{tabular}
\end{center}
\end{table}

We perform fits to the first and second moments to obtain values of the heavy--quark 
parameters which are less dependent on the details of the spectral shape. 
We take into account the correlation coefficients between the errors on the 
moments.
The full correlation matrices are given in Appendix~\ref{sec:appendix}.

Using the shape function scheme, but ignoring theoretical uncertainties, 
we fit to the first and second moments at the lowest minimum photon energy $E_{\gamma}^{min}=1.897\,\gev$.
The fitted values are $m_\b=(4.60^{+0.12}_{-0.14})\gev$ and $\mu_{\pi}^2=(0.19^{+0.22}_{-0.20})\gev^2$.
These are in agreement with the fit to the full spectral shape, but have larger errors. 
A fit to the moments at the higher photon energy $E_{\gamma}^{min}=2.094\,\gev$ gives 
$m_\b=(4.53^{+0.11}_{-0.14})\gev$ and $\mu_{\pi}^2=(0.35^{+0.13}_{-0.14})\gev^2$, 
and a fit to the highest photon energy $E_{\gamma}^{min}=2.261\,\gev$ reproduces the results from  
the fit to the spectrum. 

Results with similar precision are obtained from fits 
using the kinetic scheme if theoretical uncertainties 
are ignored. At the higher photon energies the theoretical predictions for inclusive quantities 
such as moments are not very reliable, as shown by the solid lines in Figure~\ref{fig:moments}.
It appears that an optimal combination of experimental and theoretical uncertainties occurs 
in the range $E_{\gamma}^{min}=1.9-2.1\,\gev$. 

\section{Conclusions} \label{sec:conc}

We have measured the inclusive $b\to s\gamma$ branching fraction for $E_{\gamma}>1.9\,\gev$ 
and extrapolated it to $E_{\gamma}>1.6\,\gev$. 
Our result is in good agreement with the world average.
Although our measurement is currently systematics limited,
with more data we expect to improve our understanding
of the missing fraction of high multiplicity $X_s$ hadronic final states,
and therefore reduce our dominant systematic error.
We have made the first measurement of the isospin asymmetry 
between $\Bm\to X_{s\bar{u}}\gamma$ and $\Bzb\to X_{s\bar{d}}\gamma$, 
and found that it is consistent with zero within the experimental uncertainty, 
which is mainly statistical.

We have made a measurement of the $\b\to s\gamma$ spectral shape 
over the range $E_\gamma>1.9\,\gev$.
After taking into account the presence of the $\Kstar(892)$ resonance, 
the shape is found to agree well with two recent theoretical calculations.
Fits to the spectrum are used to give values for the heavy--quark parameters 
$m_\b$ and $\mu_{\pi}^2$ in the two schemes. 
We calculate the first, second and third moments of the photon spectrum 
for five different minimum values of the photon energy between 1.90 and 2.26\,\gev.
The moments are in good agreement with predictions based 
on fits to the measured $\b\to c\ell\nu$ moments. Fits to 
the photon energy moments give heavy--quark parameters 
which agree with the fits to the full spectrum, but are less accurate. 
Information from $\b\to s\gamma$ and $\b\to c\ell\nu$ 
can be combined to obtain tighter constraints on $m_\b$ and $\mu_{\pi}^2$.
This will lead to improved extractions of \Vub\ from the measurements 
of $b\to u\ell\nu$ decays.

While this paper was in preparation, new references appeared concerning both
the photon energy spectrum and the moments.
Ref.~\cite{DGE} computes the photon energy spectrum and moments by resummed
perturbation theory, using the technique of Dressed Gluon Exponentiation.
The predicted spectrum extends smoothly into the non--perturbative region and
tends to zero at the physical endpoint.
The predictions for the moments have been compared to our results
and are consistent with them.
Ref.~\cite{Limosani}
presents fits to the BELLE spectrum~\cite{belleinclbsg} in the shape function scheme.
Ref.~\cite{AdvancedNeubert} shows
predictions for the photon energy moments using an OPE approach calculated to NNLO accuracy.
This does not require spectral information from the shape
functions. These predictions are fit to the experimental moments
from the BELLE spectrum and to the moments from this analysis.
The new preprints indicate the current interest in the extraction of
heavy--quark parameters from the shape of the $b\to s\gamma$ spectrum, and show how
our data will contribute to improved knowledge of these parameters. \\

% Input the pubboard acknowledgements file
\section{Acknowledgments} % need a endline in conclusion.tex or there is no space before the heading 
We have greatly benefited from many discussions with D. Benson, I.I. Bigi, A.L. Kagan, M. Neubert and N. Uraltsev. 
In particular,  we would like to thank A.L. Kagan for providing us with a computer code implementing the calculations in Ref.~\cite{KaganNeubert},
M. Neubert for making available for us a program with the shape function scheme calculations 
and D. Benson for providing us with tabulated values of the photon energy spectrum in the kinetic scheme.
We are grateful for the 
extraordinary contributions of our \pep2\ colleagues in
achieving the excellent luminosity and machine conditions
that have made this work possible.
The success of this project also relies critically on the 
expertise and dedication of the computing organizations that 
support \babar.
The collaborating institutions wish to thank 
SLAC for its support and the kind hospitality extended to them. 
This work is supported by the
US Department of Energy
and National Science Foundation, the
Natural Sciences and Engineering Research Council (Canada),
Institute of High Energy Physics (China), the
Commissariat \`a l'Energie Atomique and
Institut National de Physique Nucl\'eaire et de Physique des Particules
(France), the
Bundesministerium f\"ur Bildung und Forschung and
Deutsche Forschungsgemeinschaft
(Germany), the
Istituto Nazionale di Fisica Nucleare (Italy),
the Foundation for Fundamental Research on Matter (The Netherlands),
the Research Council of Norway, the
Ministry of Science and Technology of the Russian Federation, and the
Particle Physics and Astronomy Research Council (United Kingdom). 
Individuals have received support from 
CONACyT (Mexico),
the A. P. Sloan Foundation, 
the Research Corporation,
and the Alexander von Humboldt Foundation.

\appendix
\clearpage

%\onecolumngrid

\section{}
\label{sec:appendix}

In this appendix we present more detailed information concerning the fit to the spectrum
and the correlation matrices for the moments.

\subsection{Ellipse from the Fit to the Spectrum for the Shape Function Scheme}

Table~\ref{tab:SFfit} lists the points on the ellipse from the fit to the 
spectrum using the exponential model in the shape function scheme
and shown in Figure~\ref{fig:ellipses}.

\begin{table}[htbp]
\caption{Heavy--quark parameters, $m_\b$ and $\mu_\pi^2$, evaluated along the
$\Delta \chi^2 = 1$ contour resulting
from the fit to the spectrum using the shape function 
scheme.}
\begin{center}
\begin{tabular}{c c | c c }
\hline\hline
$m_\b (\gev)$ & $\mu_\pi^2 (\gev^2)$  & $m_\b (\gev)$ & $\mu_\pi^2 (\gev^2)$   \\ \hline \vspace{-.1in}\\\vspace{.04in}
4.618 & 0.249 &4.730 & 0.088 \\\vspace{.04in}
4.629 & 0.240 &4.721 & 0.095 \\\vspace{.04in}
4.636 & 0.233 &4.716 & 0.099 \\\vspace{.04in}
4.642 & 0.227 &4.709 & 0.106 \\\vspace{.04in}
4.647 & 0.222 &4.702 & 0.112 \\\vspace{.04in}
4.655 & 0.213 &4.697 & 0.118 \\\vspace{.04in}
4.662 & 0.205 &4.692 & 0.124 \\\vspace{.04in}
4.669 & 0.197 &4.686 & 0.129 \\\vspace{.04in}
4.675 & 0.190 &4.681 & 0.135 \\\vspace{.04in}
4.680 & 0.183 &4.677 & 0.140 \\\vspace{.04in}
4.685 & 0.177 &4.672 & 0.146 \\\vspace{.04in}
4.690 & 0.171 &4.667 & 0.152 \\\vspace{.04in}
4.694 & 0.165 &4.659 & 0.162 \\\vspace{.04in}
4.702 & 0.154 &4.650 & 0.173 \\\vspace{.04in}
4.709 & 0.145 &4.642 & 0.183 \\\vspace{.04in}
4.715 & 0.136 &4.634 & 0.194 \\\vspace{.04in}
4.721 & 0.127 &4.627 & 0.204 \\\vspace{.04in}
4.726 & 0.120 &4.620 & 0.214 \\\vspace{.04in}
4.730 & 0.113 &4.614 & 0.224 \\\vspace{.04in}
4.734 & 0.106 &4.608 & 0.233 \\\vspace{.04in}
4.737 & 0.100 &4.604 & 0.241 \\\vspace{.04in}
4.739 & 0.095 &4.601 & 0.249 \\\vspace{.04in}
4.741 & 0.090 &4.601 & 0.255 \\\vspace{.04in}
4.742 & 0.087 &4.603 & 0.257 \\\vspace{.04in}
4.740 & 0.084 &4.607 &  0.257 \\
\hline\hline
\end{tabular}
\end{center}
\label{tab:SFfit}
\end{table}

\subsection{Correlation Matrices for the Moments}

Here we present the correlation matrices for the first and second
moments. Table~\ref{tab:correl_stat} shows the statistical correlation matrix
and Table~\ref{tab:correl_syst} shows the systematic correlation matrix.

\begin{table}[htbp]
\begin{center}
\caption[]{\label{tab:correl_stat}{Statistical correlation coefficients for the moments
with different minimum cuts on the photon energy.}}
\end{center}
\begin{tabular}{ccccccc}
\hline\hline
& $E^{min}$(\gev) & 1.897 &  1.999  & 2.094 & 2.181 &  2.261 \\\hline \vspace{-.1in}\\\vspace{.04in}
                  &       & \multicolumn{5}{c}{$\langle E\rangle$} \\ \hline  \vspace{-.1in}\\\vspace{.04in}
                  & 1.897 &    1    & 0.55  & 0.28  &  0.12  & 0.033 \\\vspace{.04in}
                  & 1.999 &         &  1    & 0.47  &  0.19  & 0.046 \\\vspace{.04in}
$\langle E\rangle$& 2.094 &         &       & 1     &  0.48  & 0.18 \\\vspace{.04in}
                  & 2.181 &         &       &       &    1   & 0.45 \\\vspace{.04in}
                  & 2.261 &         &       &       &        & 1 \\
\hline \vspace{-.1in}\\\vspace{.04in}
                                          &       & \multicolumn{5}{c}{$\langle(E-\langle E\rangle)^2\rangle$} \\\hline \vspace{-.1in}\\\vspace{.04in}
                                          & 1.897 & 1       &  0.28 & 0.03  & -0.07  & -0.06  \\\vspace{.04in}
                                          & 1.999 &         &   1   & 0.16  & -0.14  & -0.14   \\\vspace{.04in}
 $\langle(E-\langle E\rangle)^2\rangle$   & 2.094 &         &       &   1   &  0.13  & -0.09   \\\vspace{.04in}
                                          & 2.181 &         &       &       &   1    & 0.17 \\\vspace{.04in}
                                          & 2.261 &         &       &       &        & 1  \\
\hline \vspace{-.1in}\\\vspace{.04in}
                  &       & \multicolumn{5}{c}{$\langle(E-\langle E\rangle)^2\rangle$} \\\hline \vspace{-.1in}\\\vspace{.04in}
                  & 1.897 &  -0.90  & -0.41 & -0.23 & -0.12 & -0.05 \\\vspace{.04in}
                  & 1.999 &         & -0.79 & -0.40 & -0.19 & -0.08  \\\vspace{.04in}
$\langle E\rangle$& 2.094 &         &       & -0.79 & -0.42  & -0.18  \\\vspace{.04in}
                  & 2.181 &         &       &       & -0.82  & -0.37  \\\vspace{.04in}
                  & 2.261 &         &       &       &        & -0.76  \\
\hline
\hline
\end{tabular}
\end{table}

\begin{table}[htbp]
\begin{center}
\caption[]{\label{tab:correl_syst}{Systematic correlation coefficients for the moments
with different minimum cuts on the photon energy.}}
\end{center}
\begin{tabular}{ccccccc}
\hline\hline
& $E^{min}$(\gev) & 1.897 &  1.999  & 2.094 & 2.181 &  2.261 \\\hline \vspace{-.1in}\\\vspace{.04in}
                  &       & \multicolumn{5}{c}{$\langle E\rangle$} \\ \hline \vspace{-.1in}\\\vspace{.04in}
                  & 1.897 &    1    & 0.98  & 0.97  &  0.93  & 0.85 \\\vspace{.04in}
                  & 1.999 &         &  1    &  0.98 &  0.95  & 0.84 \\\vspace{.04in}
$\langle E\rangle$& 2.094 &         &       & 1     &  0.97  & 0.86 \\\vspace{.04in}
                  & 2.181 &         &       &       &    1   & 0.92 \\\vspace{.04in}
                  & 2.261 &         &       &       &        & 1 \\
\hline \vspace{-.1in}\\\vspace{.04in}
                                          &       & \multicolumn{5}{c}{$\langle(E-\langle E\rangle)^2\rangle$} \\\hline \vspace{-.1in}\\\vspace{.04in}
                                          & 1.897 & 1       &  0.87 & 0.86  &  0.62  &  0.30  \\\vspace{.04in}
                                          & 1.999 &         &   1   & 0.90  &  0.70  &  0.27   \\\vspace{.04in}
 $\langle(E-\langle E\rangle)^2\rangle$   & 2.094 &         &       &   1   &  0.85  &  0.38   \\\vspace{.04in}
                                          & 2.181 &         &       &       &   1    & 0.67 \\\vspace{.04in}
                                          & 2.261 &         &       &       &        & 1  \\
\hline \vspace{-.1in}\\\vspace{.04in}
                  &       & \multicolumn{5}{c}{$\langle(E-\langle E\rangle)^2\rangle$} \\\hline \vspace{-.1in}\\\vspace{.04in}
                  & 1.897 &  -0.93  & -0.87 & -0.91 & -0.71 & -0.26 \\\vspace{.04in}
                  & 1.999 &         & -0.90 & -0.91 & -0.74 & -0.22  \\\vspace{.04in}
$\langle E\rangle$& 2.094 &         &       & -0.90 & -0.75  & -0.21  \\\vspace{.04in}
                  & 2.181 &         &       &       & -0.73  & -0.20  \\\vspace{.04in}
                  & 2.261 &         &       &       &        & -0.08  \\
\hline
\hline
\end{tabular}
\end{table}
\clearpage


\begin{thebibliography}{99}
\bibitem{Nskfunction}
The Novosibirsk function is defined as $f(\mes) = A_S \exp{( -0.5 \{ \ln^2[ 1
  + \Lambda\tau\cdot(\mes - m_0) ]/\tau^2 +\tau^2 \} )}$, where $\Lambda =
  \sinh{( \tau \sqrt{\ln 4})}/{(\sigma \tau \sqrt{\ln 4})}$,
the peak position is $m_0$, the width is $\sigma$, and $\tau$ is the tail parameter.
\end{thebibliography}
\end{document}